\documentclass[12pt,letterpaper]{article}
\pdfoutput=1

\usepackage{graphicx,array}
\usepackage{color}
\usepackage{latexsym}
\usepackage{amsthm}
\usepackage{amsmath}
\usepackage{amssymb}
\usepackage[numbers,sort&compress]{natbib}
\usepackage{bm}
\usepackage{slashed}
\usepackage{mathrsfs}
\usepackage{hyperref} %Automatically links \label and \ref commands; Always load last
\hypersetup{
    colorlinks=true,       % false: boxed links; true: colored links
    linkcolor=red,          % color of internal links
    citecolor=blue,        % color of links to bibliography
    filecolor=magenta,      % color of file links
    urlcolor=blue           % color of external links
}
\usepackage[all]{hypcap} %Link navagates to top of figure instead of caption (below fig)

\usepackage{myterms}
\usepackage{latexsym}

\newcommand{\QN}{\text{\sc qn}}

\newcommand{\MW}{\text{\sc mw}}

\newcommand{\B}{\textsf{B}}
\renewcommand{\L}{\textsf{L}}
\newcommand{\CP}{\textsf{CP}}

\newcommand{\Hph}{\text{\rm (q)}}
\newcommand{\Lph}{\text{\rm (h)}}
\newcommand{\pvec}{{\bm p}}

\newcommand{\xvec}{\xvec}

\newcommand{\nc}{\newcommand}  

\nc{\beq}{\begin{equation}}  
\nc{\eeq}{\end{equation}}  
\nc{\beqa}{\begin{eqnarray}}  
\nc{\eeqa}{\end{eqnarray}}  
\nc{\bit}{\begin{itemize}}  
\nc{\eit}{\end{itemize}}  

\numberwithin{equation}{section}

\title{\bf Six Flavor Quark Matter}

\author{\large Yang Bai$^{a}$ and Andrew J. Long$^{b}$}
\date{\small \it 
$^a$Department of Physics, University of Wisconsin-Madison, Madison, WI 53706, USA \\
$^b$Kavli Institute for Cosmological Physics, University of Chicago, Chicago, Illinois 60637, USA 
}

\begin{document}

\maketitle

\setlength{\parskip}{0.2ex}

\begin{abstract}
Macroscopic nuggets of quark matter were proposed several decades ago as a candidate for dark matter.  The formation of these objects in the early universe requires the QCD phase transition to be first order --- a requirement that is not satisfied in the Standard Model where lattice simulations reveal a continuous crossover instead.  In this article we point out that new physics may supercool the electroweak phase transition to below the QCD scale, and the QCD phase transition with six massless quarks becomes first-order.  As a result, the quark nuggets composed of six-flavor quark matter (6FQM) may survive as a viable dark matter candidate.  The size of a 6FQM nugget is estimated to be around $10^{10}$ grams in mass and $10^{-2}$ cm in radius.  The calculated relic abundance of 6FQM nuggets is comparable to the observed dark matter energy density; therefore, this scenario provides a compelling explanation for the coincident energy densities of dark and baryonic matter. We have explored various potential signatures --- including a gravitational wave background, gravitational lensing, and transient photon emission from collisions with compact stars and other nuggets --- and demonstrated that the favored region of parameter space is still allowed by current constraints while discovery of 6FQM nugget dark matter may require new experimental probes.  
\end{abstract}

\newpage

\begingroup
\hypersetup{linkcolor=black}
\tableofcontents
\endgroup

\newpage

%==================================
% Introduction
%==================================
\section{Introduction}\label{sec:Introduction}

%=========
Although the presence of dark matter in our Universe is now firmly established, the nature of dark matter remains a complete mystery.  
It has long been known that none of the Standard Model (SM) elementary particles make viable candidates for cold dark matter, since all of the stable particles are either not cold (neutrinos), or not ``dark'' (electron, proton, nuclei, atoms), or not matter (photons).  
This observation fuels the argument that new physics --- beyond the Standard Model --- is required to explain our Universe.  
In the vast majority of dark matter models, the new physics introduces a neutral, stable and weakly-interacting particle to serve as a dark matter candidate. 
Given the current null results from direct detection, indirect detection and collider-based searches for elementary dark matter particles, it is an opportune time to consider that dark matter may have a totally different nature than what is usually assumed and that the detection of dark matter may require entirely different measurements.  
One possibility is that dark matter is a macroscopic state made of only SM particles and fields, and the role of new physics is to provide a formation mechanism in the early universe. 
Perhaps the most familiar examples of macroscopic SM dark matter candidates are primordial black holes~\cite{Carr:1974nx}, although they are unlikely to account for all of the dark matter in light of various, stringent observational constraints~\cite{Carr:2017jsz}.  
In this paper, we are exploring another macroscopic dark matter candidate --- quark nuggets~\cite{Witten:1984rs}. 

%=========
The quark nugget is an exotic, macroscopic object composed of $u$, $d$, and $s$ quarks that was proposed by Witten in \rref{Witten:1984rs}, where he also suggested that quark nuggets could provide a natural dark matter candidate.  Witten argued that a first order quark-hadron phase transition in the early universe could concentrate baryon number into localized regions that survive in the universe today as quark nuggets.  The quark matter that makes up a nugget has a lower energy per baryon than a free proton or iron nucleus, implying that quark nuggets are the energetically stable ``ground states" of quantum chromodynamics (QCD). The estimated number of baryons inside a quark nugget is astronomically large; typical masses reach $10^9 - 10^{18}$ grams, and with a nuclear-scale energy density, $\sim10^{15} \gram / {\rm cm}^3$, the nugget radius is approximately $10^{-2} - 10 \cm$. Therefore, Witten's nuggets of quark matter provide a macroscopic dark matter candidate, which is reminiscent of a micro-neutron star, with a variety of interesting phenomenology and unique signatures \cite{Farhi:1985ib,Alcock:1985vc,Olinto:1986je,Madsen:1986jg,Alcock:1986hz,Alcock:1986bw,Olinto:1987bt,Alcock:1988re,Alcock:1988br,Frieman:1989bu,Olinto:1991rr,Olinto:1991qq,Olesen:1991zt,Olesen:1993iz,Madsen:1998uh,Wilczek:2004rm,Wilczek:2005ez,Madsen:2009tb,Han:2009sj,Lawson:2012zu}. 

%=========
One of the key assumptions in Witten's argument is that the QCD phase transition is a first-order one.  
Unfortunately, numerical lattice studies~\cite{Fodor:2001pe} have demonstrated that the SM quark-hadron transition is a continuous crossover rather than a first order phase transition, implying that SM physics alone is unable to form quark nuggets in the early universe.  
In this article we explore the idea that new physics can affect the order of the QCD phase transition and resuscitate quark nugget dark matter.  

 %=========
For a QCD-like gauge theory, it is well known~\cite{Pisarski:1983ms,Brown:1990ev} that the phase transition would be first order if the number of light quarks below the confinement scale, $N_f$, were greater than or equal to $N_c=3$.  (This is not the case in the SM, because of the medium-heavy strange quark mass.)  Clearly the SM cannot be extended to include new light quarks, which would lead to new, undiscovered hadron states and affect the running of the strong coupling.  However, the SM contains a total of six quark flavors, and although half of them are heavy today ($m_{c, b, t} > \Lambda_\QCD$), it is possible that their masses in the early universe could have been very different.  If at least three quark flavors are massless at the QCD epoch, the QCD phase transition is expected to be a first-order one, and Witten's argument --- with some modifications that we discuss --- implies the formation of quark nuggets.  

 %=========
There may exist many scenarios in which the quarks are massless at the QCD epoch.  For instance, some quarks may have their masses proportional to the vacuum expectation value (VEV) of a flavon field, which only reaches the minimum of its potential much later.  However, our primary interest here is with another possibility; namely, that electroweak symmetry breaking is delayed until the QCD epoch due to the influence of some new physics~\cite{Kuzmin:1992up,Konstandin:2011dr,Servant:2014bla,Arunasalam:2017ajm,Iso:2017uuu,vonHarling:2017yew} (see also \rref{Quigg:2009xr}). Then all six flavors of SM quarks are massless during the QCD phase transition, which is expected to be first order, leading to the formation of quark nuggets that contain all six quark flavors.  

%=========
In this article we argue that a supercooled electroweak phase transition can trigger a first order quark-hadron transition, which results in the formation of nuggets containing six-flavor quark matter (6FQM).  
We consider 6FQM nuggets as a dark matter candidate.  
One should take care to distinguish our work from Refs.~\cite{Farrar:2002ic,Farrar:2017eqq,Gross:2018ivp}, in which it is proposed that the dark matter is a 6-quark hadron (sexaquark) with quark content $uuddss$.  
An alternative idea for generating quark nuggets with the aide of an axion field has been proposed in \rref{Zhitnitsky:2002qa} and explored further in Refs.~\cite{Oaknin:2003uv,Lawson:2013bya,Liang:2016tqc,Ge:2017idw}.  

%=========
The remainder of this article is organized as follows.  
In \sref{sec:SFQM} we invite the reader to consider a phase of SM matter, which we call six-flavor quark matter, that has a vanishing Higgs VEV, vanishing QCD condensates, six flavors of massless quarks, and nonzero baryon and lepton numbers.  
In \sref{sec:Production} we explain how nuggets of six-flavor quark matter could have been formed in the early universe, and in \sref{sec:Stability} we argue that they are cosmologically stable.  
We discuss various phenomenological implications and observational signatures of these quark nuggets in \sref{sec:Phenomenology}, and we conclude the article in \sref{sec:Conclusion}.  

%==================================
% Six Flavor Quark Matter
%==================================
\section{Six Flavor Quark Matter}
\label{sec:SFQM}

%=========
Many-body systems at finite temperature and density can exhibit a rich phase structure.  
Here we are interested in a particular phase of the Standard Model particles that arises when 1) the expectation value of the Higgs field vanishes, 2) the quark and gluon condensates vanish, 3) the baryon and lepton numbers are nonzero, 4) the gauge charges are zero, and 5) the temperature is low.  
This phase shares similar properties to Witten's quark matter phase \cite{Witten:1984rs}, but since the Higgs field vanishes and all six quarks species are massless, we refer to our phase as six-flavor quark matter (6FQM).  
The assumption of a vanishing Higgs vacuum expectation value is reminiscent of the {\it Gedanken} world studied in \rref{Quigg:2009xr}.  
However, we will argue in \sref{sec:Production} that 6FQM it not merely hypothetical, but rather this exotic phase may exist in localized regions of our Universe.  

%=========
In the phase of six-flavor quark matter, the electroweak symmetry is unbroken and color is not confined.  
Thus we enumerate the SM particles as multiplets under the $\SU{3}_c \times \SU{2}_L \times \U{1}_Y$ gauge symmetry and the $\U{1}_\B \times \U{1}_{\L_1} \times \U{1}_{\L_2} \times \U{1}_{\L_3}$ global symmetry: 
\begin{align}\label{eq:charges}
\renewcommand{\arraystretch}{1.1}
\begin{array}{c||c|c|c||c|c|c|c}
	& \SU{3}_c & \SU{2}_L & \U{1}_Y & \U{1}_\B & \U{1}_{\L_1} & \U{1}_{\L_2} & \U{1}_{\L_3} \\ 
	\hline \hline 
	q_L^i & {\bm 3} & {\bm 2} & 1/6 & 1/3 & 0 & 0 & 0 \\ 
%	\bar{q}_L^i & \bar{\bm 3} & \bar{\bm 2} & -1/6 & -1/3 & 0 & 0 & 0 \\ 
%	\hline
	u_R^i & {\bm 3} & {\bm 1} & 2/3 & 1/3 & 0 & 0 & 0 \\ 
%	\bar{u}_R^i & \bar{\bm 3} & {\bm 1} & -2/3 & -1/3 & 0 & 0 & 0 \\ 
%	\hline
	d_R^i & {\bm 3} & {\bm 1} & -1/3 & 1/3 & 0 & 0 & 0 \\ 
%	\bar{d}_R^i & \bar{\bm 3} & {\bm 1} & 1/3 & -1/3 & 0 & 0 & 0 \\ 
%	\hline
	l_L^i & {\bm 1} & {\bm 2} & -1/2 & 0 & \delta_{i1} & \delta_{i2} & \delta_{i3} \\ 
%	\bar{l}_L^i & {\bm 1} & \bar{\bm 2} & 1/2 & 0 & -\delta_{i1} & -\delta_{i2} & -\delta_{i3} \\ 
%	\hline
	e_R^i & {\bm 1} & {\bm 1} & -1 & 0 & \delta_{i1} & \delta_{i2} & \delta_{i3} \\ 
%	\bar{e}_R^i & {\bm 1} & {\bm 1} & 1 & 0 & -\delta_{i1} & -\delta_{i2} & -\delta_{i3} \\ 
%	\hline
	\Phi & {\bm 1} & {\bm 2} & 1/2 & 0 & 0 & 0 & 0 \\ 
%	\bar{\Phi} & {\bm 1} & \bar{\bm 2} & -1/2 & 0 & 0 & 0 & 0 \\ 
	\hline 
	G & {\bm 8} & {\bm 1} & 0 & 0 & 0 & 0 & 0 \\ 
	W & {\bm 1} & {\bm 3} & 0 & 0 & 0 & 0 & 0 \\ 
	Y & {\bm 1} & {\bm 1} & 0 & 0 & 0 & 0 & 0 \\
\end{array} \per
\end{align}
Antiparticles to the fermions and Higgs bosons are denoted with a bar (not shown).  
Gauge indices are suppressed, and the flavor index $i \in \{ 1, 2, 3 \}$ labels the generation.  
One can extend the Standard Model to explain neutrino masses and the phenomenon of neutrino flavor oscillations, and the rest of our analysis is unchanged as long as the new interactions do not come into chemical equilibrium at $T \lesssim 1 \GeV$.\footnote{If the neutrinos are Dirac particles that acquire their masses from a tiny Yukawa coupling, $y_\nu \sim m_\nu / v_\EW \sim 10^{-12}$, then the Yukawa interaction does not come into equilibrium at any time.  If the neutrinos are Majorana particles that acquire their masses from the seesaw mechanism, then the lepton-number-violating interactions are out of equilibrium at temperatures below $T \sim \Lambda^2 / \Mpl \sim 10^{10} \GeV$ where $\Lambda^{-1} \sim m_\nu / v_\EW^2$ is the coefficient of the dimension-5 Weinberg operator.}

%=========
The Standard Model particles scatter though gauge, Yukawa and sphaleron interactions:\footnote{Recall that for massless fermions, we can work in a flavor basis where the gauge interactions, the down-type quark Yukawa interactions, and the electron Yukawa interactions are flavor diagonal, but the up-type quark Yukawa interactions are flavor changing.  Thermal effects will lift the fermion masses and select a preferred basis.  The universal gauge interactions lead to a degenerate spectrum, but the Yukawa interactions lead to a splitting.  }
\begin{align}\label{eq:reactions}
\renewcommand{\arraystretch}{1.1}
\begin{array}{c|l}
	\text{interaction} & \text{reaction} \\ \hline
	\text{strong gauge} & x + \bar{x} \longleftrightarrow G, \, 2G, \cdots \quad \text{for} \quad x \in \{ q_L^i, u_R^j, d_R^k \} \\ 
	\text{weak gauge} & x + \bar{x} \longleftrightarrow W, \, 2W, \cdots \quad \text{for} \quad x \in \{ q_L^i, l_L^j \} \\ 
	\text{hypercharge} & x + \bar{x} \longleftrightarrow Y, \, 2Y, \cdots \quad \text{for} \quad x \in \{ q_L^i, u_R^i, d_R^i, l_L^i, e_R^i, \Phi \} \\ 
	\text{up-type quark Yukawa} & q_L^i + \bar{u}_R^j \longleftrightarrow \bar{\Phi} \\ 
	\text{down-type quark Yukawa} & q_L^i + \bar{d}_R^i \longleftrightarrow \Phi \\ 
	\text{charged-lepton Yukawa} & l_L^i + \bar{e}_R^i \longleftrightarrow \Phi \\
	\text{weak sphaleron} & \sum_{i} \, \bigl( q_L^i + q_L^i + q_L^i + l_L^i \bigr) \longleftrightarrow 0 \\
	\text{strong sphaleron} & \sum_{i} \, \bigl( q_L^i + q_L^i + \bar{u}_R^i + \bar{d}_R^i \bigr) \longleftrightarrow 0 
\end{array} 
	\com
\end{align}
where color and weak isospin indices are suppressed.  
Note that the gauge interactions may involve a single gauge boson or several.  

%=========
Inspecting the reactions in \eref{eq:reactions} reveals that there are three conserved global charges: $\B/3-\L_i$ for $i \in \{ 1,2,3 \}$ where $\B$ is baryon number and $\L_i$ is lepton number of the $i^{\rm th}$ generation.  
(The gauge charges are also conserved and assumed to vanish.)  
We are interested in a system with nonzero values for the corresponding conserved charge densities, $n_{\B/3-\L_i}$.  
As this system cools to a critical temperature where $T^3 \sim n_{\B/3-\L_i}$, the fermions form a degenerate Fermi gas.  
This is the phase of six-flavor quark matter.  

%=========
In following analysis we will assume that the Higgs bosons in the electroweak-unbroken phase acquire a (small) mass, $m_\Phi \sim 1 - 10 \GeV$.  
This implies a convex effective potential, $V_{\rm eff}^{\prime\prime}(h=0) = m_\Phi^2 > 0$, that helps to stabilize the electroweak-unbroken phase and leads to the desired supercooling of the electroweak phase transition; see also \sref{sec:EW_Trans}.  
When the system cools to a temperature, $T < m_\Phi$, the three-body Yukawa interactions in \eref{eq:reactions} go out of equilibrium, because reactions producing an on-shell Higgs boson are Boltzmann suppressed.  
The Higgs and anti-Higgs bosons in the system decay to quark and lepton pairs, which transfers any particle-antiparticle asymmetry carried by these species to the fermions.  
Thus, to describe the system at low temperatures ($T < m_\Phi$) we require only the two-to-two Yukawa interactions mediated by an off-shell Higgs boson.  

%=========
Now we seek to determine the flavor content of six-flavor quark matter and to calculate its energy density.  
For each particle species we assign a chemical potential: $\mu_{q_L^i}, \mu_{u_R^i},  \mu_{d_R^i}, \cdots, \mu_{W}, \mu_{Y}$.  
In principle one can construct a system of Boltzmann equations, and solve for the evolution of the various chemical potentials as the quark matter phase cools.  
However, all of the reactions are in equilibrium\footnote{We only require that the reactions remain in equilibrium at the temperatures when the quark matter is forming.  If the QCD phase transition occurs at $T \sim 100 \MeV$, then it is reasonable to expect that the quark matter formation is completed by $T \sim 1 \MeV$.  Specifically, we assume that the non-perturbative sphaleron reactions remain in equilibrium at these temperatures, since the system is in the electroweak-unbroken phase~\cite{Arnold:1996dy}. } at $T \sim \Lambda_\QCD$ in the quark matter phase.  
Since the gauge interactions are in equilibrium, the matter particles related by $\CP$-conjugation have opposite chemical potentials, $\mu_{\bar{x}} = - \mu_{x}$, and the self-adjoint gauge bosons have vanishing chemical potentials, $\mu_G = \mu_W = \mu_Y = 0$.  
This leaves the fifteen variables: $\{ \mu_{q_L^i}, \mu_{u_R^i}, \mu_{d_R^i}, \mu_{l_L^i}, \mu_{e_R^i} \}$.  
From the equilibrium reactions we infer a system of chemical equilibrium conditions, which are solved by
\begin{subequations}\label{eq:mu-constraint}
\begin{align}
	\mu_{q_L^1} = \mu_{q_L^2} = \mu_{q_L^3} & \equiv \mu_{q_L}  \com \\ 
	\mu_{u_R^1} = \mu_{u_R^2} = \mu_{u_R^3} & \equiv \mu_{u_R}  \com \\ 
	\mu_{d_R^1} = \mu_{d_R^2} = \mu_{d_R^3} & \equiv \mu_{d_R} = 2 \mu_{q_L} - \mu_{u_R} \com \\ 
	\mu_{l_L^1} + \mu_{l_L^2} + \mu_{l_L^3} + 9 \mu_{q_L} & =  0   \com \\ 
	\mu_{l_L^1} - \mu_{e_R^1} = \mu_{l_L^2} - \mu_{e_R^2} = \mu_{l_L^3} - \mu_{e_R^3} & = \mu_{u_R} - \mu_{q_L} 
	\per 
\end{align}
\end{subequations}
The chemical potential $\mu_\Phi$ is absent, because we have assumed $T < m_\Phi$ and the Higgs boson abundance is Boltzmann suppressed.  

%=========
Once the system has cooled to $T \ll \mu$, it can be described as a degenerate Fermi gas.  
Then the number density of $f$-number ({\it i.e.}, the number density of $f$ minus the number density of $\bar{f}$) is given by $n_f = g_f \, \mu_f^3/(6\pi^2)$ for a species with chemical potential $\mu_f$ and where $g_f$ counts the internal degrees of freedom ($g_{q_L^i} = 6, g_{u_R^i} = g_{d_R^i} = 3, g_{l_L^i} = 2, g_{e_R^i} = 1$).  
The density of hypercharge and the three conserved global charges are written as
\begin{subequations}\label{eq:conserved}
\begin{align}
	n_Y & = \sum_{i=1,2,3} \Bigl( \frac{1}{6} \, n_{q_L^i} \,+ \,\frac{2}{3}\,n_{u_R^i}\,-\,\frac{1}{3} \, n_{d_R^i}\,-\, \frac{1}{2} \, n_{l_L^i} \, - \, n_{e_R^i} \Bigr) \com \\ 
	n_{\B/3-\L_i} & = \frac{1}{3} \sum_{j=1,2,3} \Bigl( \frac{1}{3} n_{q_L^j} + \frac{1}{3} n_{u_R^j} + \frac{1}{3} n_{d_R^j} \Bigr) - \Bigl( n_{l_L^i} + n_{e_R^i} \Bigr) 
	\per
\end{align}
\end{subequations}
We are interested in charge-neutral systems, and we now set $n_Y = 0$.  
Moreover, we simplify by taking $n_{\B/3-\L_i} = n_{\B-\L} / 3$, which assumes that the $\B-\L$ asymmetry is uniformly distributed across the three generations.  
Now we can solve the equilibrium conditions \pref{eq:mu-constraint} with the conservation laws \pref{eq:conserved} to express all fifteen chemical potentials in terms of just $n_{\B-\L}$.  
Doing so gives
\begin{align}\label{eq:mu_sol}
	\mu_{q_L^i} = \mu 
	\ , \quad 
	\mu_{u_R^i} = -\mu 
	\ , \quad 
	\mu_{d_R^i} = 3\mu 
	\ , \quad
	\mu_{l_L^i} = -3\mu 
	\ , \quad \text{and} \quad
	\mu_{e_R^i} = -\mu  \ ,
\end{align}
where
\begin{align}\label{eq:n_sol}
	n_{\B} = \frac{14 \mu^3}{\pi^2}
	\ , \quad
	n_{\L} = -\frac{55 \mu^3}{2\pi^2}
	\ , \quad \text{and} \quad
	n_{\B-\L} = \frac{83 \mu^3}{2\pi^2}
	\per
\end{align}
The nonzero value of $\mu$ is a consequence of the nonzero $\B-\L$ charge in the quark matter.  

%=========
Until this point we have imagined a system with fixed $n_{\B-\L}$ that is externally predetermined, but now we must consider how the quark matter phase arises in a cosmological environment.  
As we will discuss further in \sref{sec:Production}, the quark matter is expected to form if the cosmological quark-hadron phase transition is a first-order one.  
After the transition, the universe is predominantly filled by the hadronic phase with some pockets of quark matter phase.  
In this setting the number density $n_{\B-\L}$ changes as the volume of the quark matter phase shrinks.  
The system approaches an equilibrium where the degeneracy pressure in the quark matter phase balances the vacuum pressure in the hadronic phase.  
At this point, we say that the pocket of quark matter phase has formed a quark nugget.  
We now determine the value of $\mu$ at the pressure equilibrium.  

%=========
The pressure difference at the phase boundary is given by 
\begin{align}\label{eq:Delta_P}
	\Delta P = \sum_{i = 1,2,3} \Bigl( P_{q_L^i} + P_{u_R^i} + P_{d_R^i} + P_{l_L^i} + P_{e_R^i} \Bigr) - B  \com 
\end{align}
where $P_f = g_f \mu_f^4/(24\pi^2)$ is the degeneracy pressure arising from massless species $f$ in the quark matter phase, and we assume that the pressure arising from particles in the hadronic phase is negligible.  
The parameter $B > 0$ is the differential vacuum pressure arising from the quark and Higgs condensates that are present in the hadronic phase, but absent in the quark matter phase.  
In the MIT bag model for hadrons, the variable $B$ is known as the bag parameter, and it takes a value $B \simeq (150 \MeV)^4$.  
(For more discussion of $B$ see \sref{sec:Stability}.)  
By imposing $\Delta P = 0$ we find the equilibrium value of $n_{\B-\L}$.  
The solution with a positive $\mu$ (equivalently, positive $n_\B$) has
\begin{align}\label{eq:equilibrium}
	\mu = \left( \frac{8 \pi^2 B}{415} \right)^{1/4} \simeq 0.66 B^{1/4} 
	, \quad 
	n_{\B} \simeq 0.41 B^{3/4} 
	, \quad 
	n_{\L} \simeq -0.80 B^{3/4} 
	, \quad 
	n_{\B-\L} \simeq 1.21 B^{3/4}
	\per
\end{align}

%=========
\begin{figure}[th!]
\begin{center}
\includegraphics[width=0.75\textwidth]{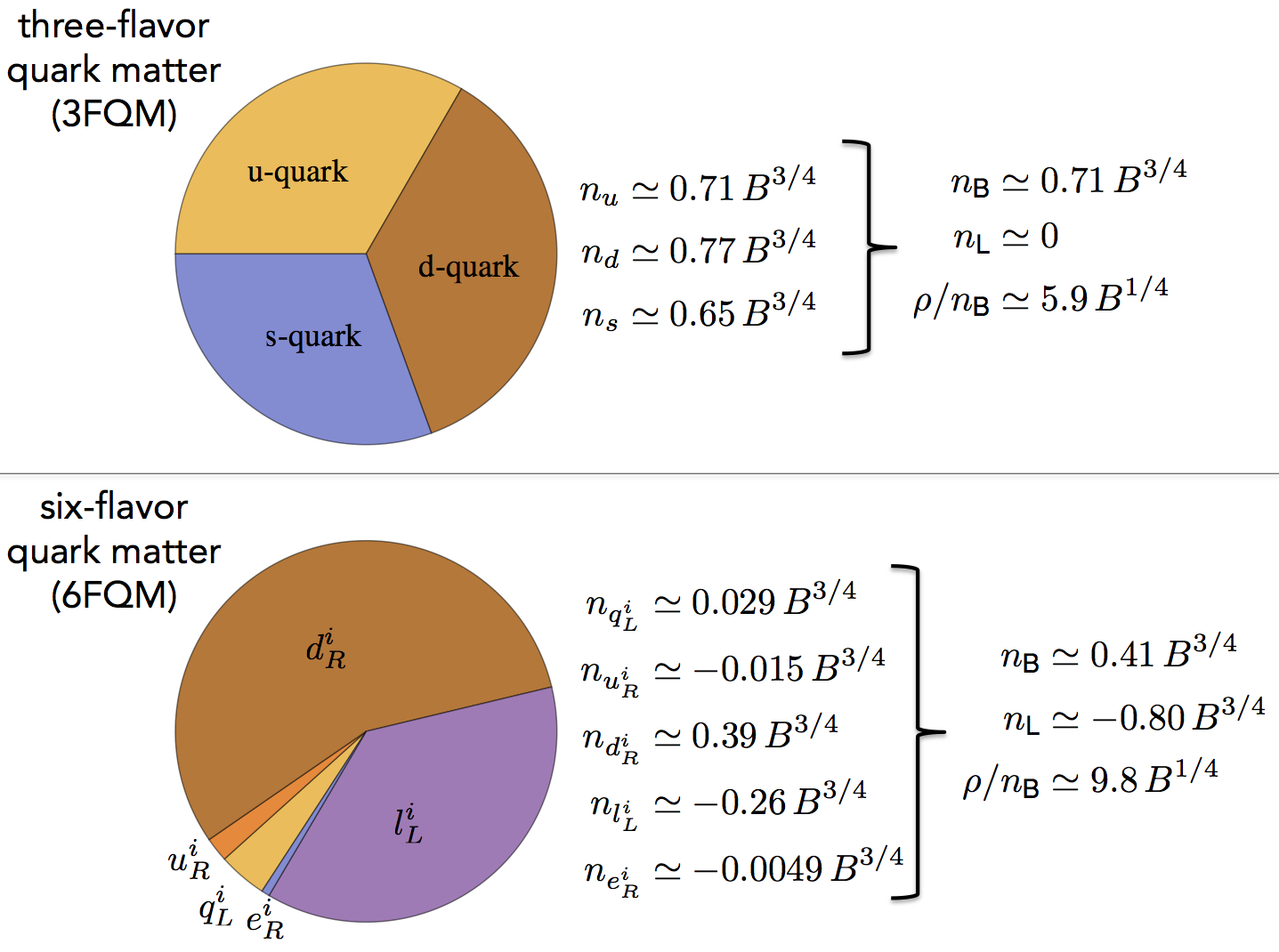} 
\caption{
\label{fig:pie_chart}
The flavor content of 6FQM that we calculated in \sref{sec:SFQM} and of 3FQM that was worked out in \rref{Farhi:1984qu}.  The charge densities, denoted by $n_f$, equal the number density of particles $f$ minus the density of $\CP$-conjugate antiparticles $\bar{f}$.  For quark densities, there is an implicit sum over colors.  In reproducing the 3FQM calculation, we have taken the strange quark mass to be $m_s \simeq 96 \MeV$, and therefore charge neutrality is obtained with approximately equal abundances of $u$, $d$, and $s$ quarks and a negligible density of electrons.  For 3FQM the differential vacuum pressure takes a value of roughly $B \simeq (150 \MeV)^4$, and we expect $B$ to be comparable for 6FQM (see \sref{sec:Stability}).  
}
\end{center}
\end{figure}

%=========
Finally let us calculate the energy-per-baryon, which is important for us to understand the mass and the stability of quark nuggets.  
The energy density in the quark matter phase is 
\begin{align}\label{eq:rho_def}
	\rho = \sum_{i = 1,2,3} \Bigl( \rho_{q_L^i} + \rho_{u_R^i} + \rho_{d_R^i} + \rho_{l_L^i} + \rho_{e_R^i} \Bigr) + B 
	\com
\end{align} 
where $\rho_f = g_f \, \mu_f^4 / (8\pi^2)$ is the energy density carried by species $f$.  
When $\mu$ reaches its equilibrium value in \eref{eq:equilibrium} we have 
\begin{align}\label{eq:rho_eq}
	\rho = 4 B 
	\per 
\end{align}
Taking the ratio of $\rho$ from \eref{eq:rho_eq} and $n_\B$ from \eref{eq:equilibrium} we obtain
\begin{align}\label{eq:rho_over_n}
	\frac{\rho}{n_\B} \simeq 9.8 B^{1/4}
	\per 
\end{align}
Additionally, if we know that the quark nugget carries $N_\B$ units of baryon number, then we determine its total energy to be 
\begin{align}\label{eq:M_QN}
	M_\QN = \frac{\rho}{n_\B} N_\B \simeq 9.8 B^{1/4} N_\B 
	\com
\end{align}
which is effectively the mass of the quark nugget.  
As we will see in \sref{sec:size_mass}, the typical mass is $M_\QN \sim 10^{10} \gram$ for $N_\B \sim 10^{34}$.  

%=========
Broadly speaking the properties of six-flavor quark matter are quite similar to those for Witten's three-flavor quark matter (3FQM).  
In both systems the differential vacuum pressure, $B$, sets the scale for the charge densities.  
However, the flavor composition is markedly different between 3FQM and 6FQM.  
We highlight this point in \fref{fig:pie_chart}, which summaries the results of this section and contrasts the properties of 3FQM and 6FQM.  

%=========
In this section we have described a phase of QCD that exists at zero temperature, but nevertheless color is not confined~\cite{Witten:1984rs}.  
At first glance this seems to be a contradiction, but it is useful to remember that the quark matter is a degenerate Fermi gas at finite density.  
It is the Fermi momentum, $p_F \sim \mu$, that sets the typical momentum transfer when two quarks scatter, and $p_F$ may be larger than the confinement scale even if the temperature is low.  
Additionally, in this finite-density system we cannot directly apply the usual calculation of the confinement scale, which asks at what energy scale does the running of the strong coupling diverge.  

%==================================
% Production in the Early Universe
%==================================
\section{Production in the Early Universe}
\label{sec:Production}

%=========
In this section we discuss how nuggets of 6FQM could be formed in the early universe during the quark-hadron phase transition.  
Since the expectation value of the Higgs field is zero in the 6FQM phase, the cosmological creation of 6FQM requires the electroweak phase transition to be supercooled below the temperature of the quark-hadron transition, which will be assumed to be true in this section.  

%-------------------------------------------
% Overview
%-------------------------------------------
\subsection{Overview}
\label{sec:overview}

\begin{figure}[t]
\begin{center}
\includegraphics[width=0.97\textwidth]{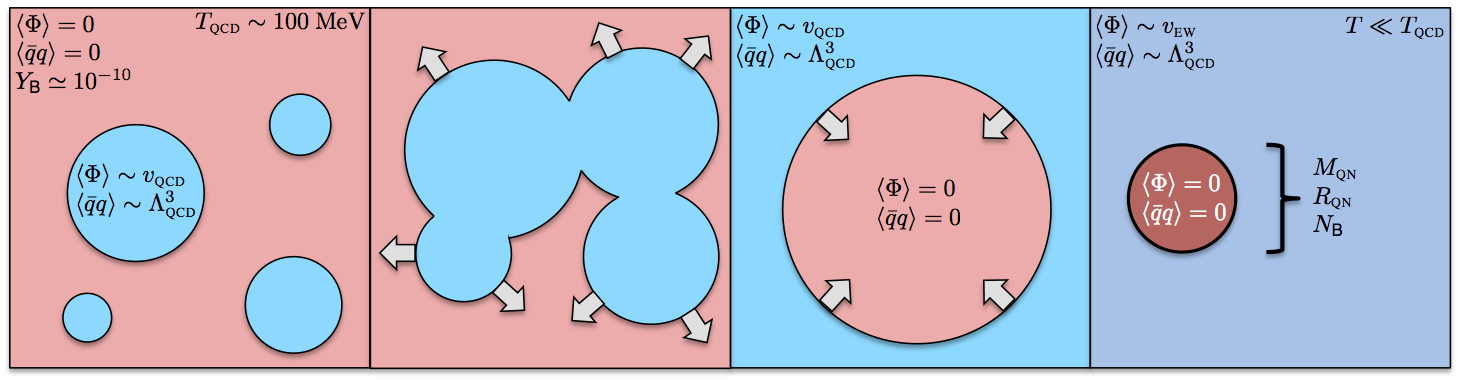}
\caption{
\label{fig:PT_cartoon}
A cartoon illustrating the cosmological dynamics leading to the formation of nuggets of six-flavor quark matter.  A first-order QCD phase transition causes the baryon number to accumulate into pockets of quark gluon plasma, which eventually cool to form 6FQM nuggets.  
} 
\end{center}
\end{figure}

%=========
In this section we briefly review the dynamics of the quark-hadron phase transition in our scenario, and discuss the formation of the quark nuggets.  
Additional details are provided in the following sections.  
The important events are illustrated in \fref{fig:PT_cartoon} and enumerated as follows:
\begin{enumerate}
	\item  The cosmological plasma is initially hotter than $T \sim 100 \GeV$, and neither the Higgs nor the chiral quark condensate have formed:  $\langle \Phi \rangle = 0$ and $\langle \bar{q} q \rangle = 0$.  Baryogenesis has already taken place, and the cosmological plasma has an excess of matter over antimatter.  
	\item  The presence of new physics in the Higgs sector supercools the electroweak phase transition to a temperature $T_\QCD \sim 100 \MeV$.  During the period of supercooling, the plasma contains six flavors of massless quarks and leptons, as well as QCD and electroweak gauge bosons.  
	\item  As the plasma temperature reaches $T_\QCD \sim 100 \MeV$, bubbles of hadronic phase begin to nucleate.  In the hadronic phase, the quark condensate takes on a nonzero value, $\langle \bar{q} q \rangle \bigr|_{T_\QCD} \sim \Lambda_\QCD^3$.  The quark condensate induces a tadpole term in the Higgs potential, and the Higgs condensate develops a nonzero value $|\langle \Phi \rangle| \sim v_\QCD$ in the hadronic phase.  
	\item  The bubbles of hadronic phase grow due to the pressure difference across the phase boundary, which is controlled by the differential vacuum pressure, $\Delta P = B$.  Particles in the plasma scatter from the expanding bubble walls, and the baryon number in the two phases becomes unequal.  The scattering induces a drag force on the bubble walls, and the bubbles expand slowly, preceded by a shock front.  
	\item  The bubbles of hadronic phase collide with one another and coalesce.  Near the end of the phase transition, the hadronic phase fills most of the Hubble volume, leaving isolated regions of quark phase with $\langle \Phi \rangle = 0$ and $\langle \bar{q} q \rangle = 0$ that will become the quark nuggets.  
	\item  After the electroweak phase transition is completed, the latent heat of the phase transition will be transferred to the plasma.  The plasma can reheat to a temperature as high as $T_{\rm rh} \sim V_0^{1/4}$ where $V_0$ is the characteristic vacuum energy difference of the metastable vacuum with $\langle \Phi \rangle \sim v_\QCD$ and $\langle \Phi \rangle \sim v_\EW$.  If $T_{\rm rh} \gtrsim T_\QCD$, then the quark condensate is ``melted,'' and the pockets of quark phase are destroyed.  To prevent this from happening, it is necessary that $V_0 \lesssim (100 \MeV)^4$ is not too large.  
	\item  We assume that the quark matter regions maintain thermal equilibrium with the ambient plasma.  As these regions cool, their pressure drops and they shrink.  Eventually the thermal pressure becomes negligible, and the pockets of quark matter, now known as quark nuggets, are supported by Fermi degeneracy pressure due to their baryon number.  
	\item  Particles scatter from the quark nuggets, which deposits heat and tends to evaporate them.  Sufficiently large nuggets can be (meta-)stable enough to survive until today.  
	\item  Quark nuggets surviving in the universe today provide a candidate for dark matter.  
\end{enumerate}

%=========

%-------------------------------------------
% Size and mass estimates
%-------------------------------------------
\subsection{Size and mass estimates}\label{sec:size_mass}

%=========
Let us estimate the size and mass of the relic quark nuggets.  
It is useful to recall that the quark nuggets formed at the QCD epoch ($t = t_\QCD$) when the temperature of the primordial plasma was roughy $T_\QCD \approx 130 \MeV$.  
If the electroweak phase transition is delayed, as we discussed in \sref{sec:overview}, then the plasma consists of $g_\ast \simeq 106.75$ effective degrees of freedom.  
Assuming that the universe is radiation dominated, the Hubble parameter, $H_\QCD$, is given by $3 \Mpl^2 H_\QCD^2 = (\pi^2/30) g_\ast T_\QCD^4$ where $\Mpl \simeq 2.43 \times 10^{18} \GeV$ is the reduced Planck mass.  
The Hubble time is given by $t_\QCD = 1 / (2 H_\QCD)$, and the Hubble radius (particle horizon) is given by $d_\QCD = 1 / H_\QCD$.  
Numerically we estimate,
\begin{align}\label{eq:QCD}
	H_\QCD & \simeq \bigl( 24 \, {\rm peV} \bigr) \, g_{106}^{1/2} \, T_{130}^{2}
	\ , \ \quad
	t_\QCD \simeq (14 \, \mu{\rm sec}) \, g_{106}^{-1/2} \, T_{130}^{-2}
	\ , \ \quad 
	d_\QCD \simeq ( 8.3 \km ) \, g_{106}^{-1/2} \, T_{130}^{-2} \,, 
\end{align}
where ${\rm peV} \equiv 10^{-12} \eV$, $g_{106} \equiv g_\ast / 106.75$ and $T_{130} \equiv T_\QCD / 130 \MeV$.  

%=========
Consider first the expanding bubbles of hadronic phase that nucleate and grow during the first-order QCD phase transition.  
The dynamics of these bubbles and their interactions with the plasma have been studied extensively by Refs.~\cite{Hogan:1984hx,Kajantie:1986hq}.  
The bubble walls are preceded by a shock front that expands with a speed $v_{\rm sh}$.  
At the time when the shock fronts collide, the hadronic phase bubbles have a typical radius given by $R_i \approx t_{\rm grow} \, v_{\rm sh}$ where $t_{\rm grow}$ is the amount of time elapsed between bubble nucleation and bubble collision.  
This time depends on the rate of bubble nucleation, and the latent heat of the phase transition.  The bubble growing time is estimated to be \cite{Kajantie:1986hq}
\begin{align}\label{eq:tgrow}
	t_{\rm grow} \approx \frac{3}{2L} \sqrt{ \frac{w_0}{16 L} } \, \sqrt{ \frac{\xi-1}{4\,\xi-1} } \, t_\QCD 
	\com
\end{align}
where $p_0$ and $w_0$ are related to the bubble nucleation probability, $L = \log[ T_c^4\, t_H^4 \,p_0\, v_{\rm sh}^3 ]$ is related to the fraction of space filled by the shock fronts, and $\xi = P_q / P_h$ is the ratio of the particle-induced pressures in the quark and hadronic phases.  
Using the results of \rref{Kajantie:1986hq} we estimate  
\begin{align}\label{eq:Ri}
	R_i \simeq 
%	\bigl(15 \cm \bigr) 
	\bigl(20 \cm \bigr) 
	\left( \frac{L}{168} \right)^{-3/2} \left( \frac{w_0}{1} \right)^{1/2} \left( \frac{g_\ast}{106.75} \right)^{-1/2} \left( \frac{T_\QCD}{130 \MeV} \right)^{-2} \left( \frac{v_{\rm sh}}{1/\sqrt{3}} \right) 
	\com
\end{align}
and we have taken $\xi = 1.9$ based on the calculation in \sref{sec:B_number}. After the hadronic-phase bubbles collide, the quark phase contains isolated regions that form their own ``bubbles" with a characteristic radius of $R_i$.
It is challenging to robustly estimate the rate of bubble nucleation and to model the bubble's interaction with the plasma, and therefore the estimates that lead to \eref{eq:Ri} represents one of the largest sources of uncertainty in our calculation; it is possible that $R_i$ could be larger or smaller by an order of magnitude. 

%=========
\begin{figure}[t]
\begin{center}
\includegraphics[width=0.55\textwidth]{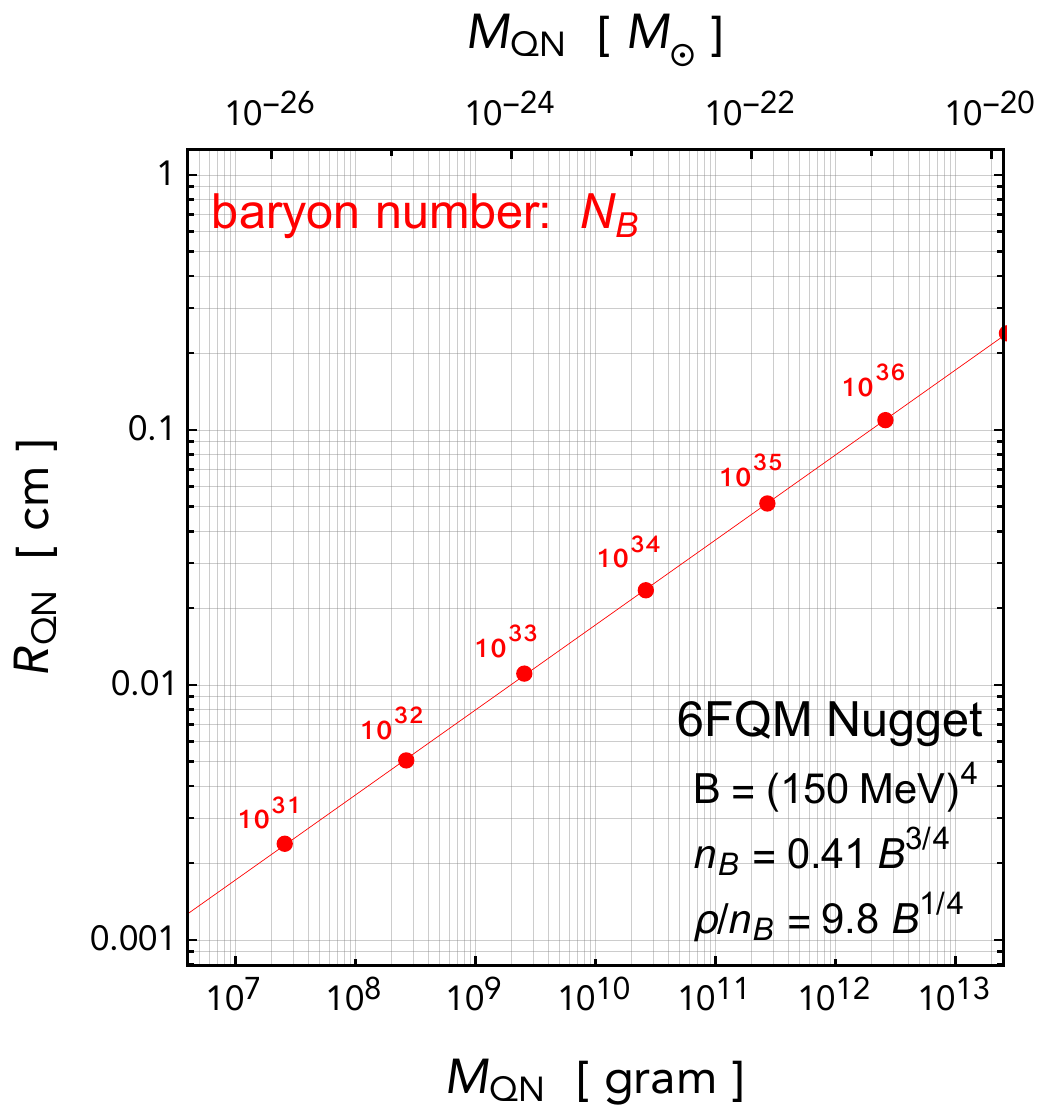}
\caption{
\label{fig:mass_radius}
The mass and radius of a nugget of 6FQM, given by \erefs{eq:R}{eq:M}, for various values of its total baryon number (shown in red).  
}
\end{center}
\end{figure}

%=========
Next we estimate the total baryon number inside of the quark nuggets.  
At the QCD epoch the cosmological density of baryon number is $n_\B = Y_\B s$ where $Y_\B \simeq 10^{-10}$ is the baryon asymmetry, and $s = (2\pi^2/45) g_{\ast S} T^3$ is the entropy density.  
Let $n_\B^\Hph$ be the density of baryon number in the pockets of quark phase.  
Thus the total baryon number inside a pocket of quark phase is $N_\B = n_\B^\Hph \, (4\pi/3) R_i^3$, which evaluates to 
\begin{align}\label{eq:NB}
	N_\B & \simeq 
%	\bigl(5.6 \times 10^{33} \bigr) 
	\bigl(6 \times 10^{33} \bigr) 
	\left( \frac{n_\B^\Hph}{Y_B s} \right) \left( \frac{Y_\B}{10^{-10}}\right) \left( \frac{g_{\ast S}}{106.75} \right) \left(\frac{T_\QCD}{130 \MeV} \right)^{3} \left( \frac{R_i}{10 \cm} \right)^{3}
	\per
\end{align}
Note that an $O(10)$ uncertainty in $R_i$ becomes an $O(10^3)$ uncertainty in $N_\B$.  

%=========
The nugget of degenerate quark matter is formed once the pocket of unbroken phase has cooled sufficiently.  
For the equilibrium quark nugget configuration that we discussed in \sref{sec:SFQM}, the density of baryon number is given by \eref{eq:equilibrium} where we estimated $n_\B \simeq 0.41 \, B^{3/4}$.   
If the total baryon number is $N_\B$ then the radius of the quark nugget can be estimated from $N_\B = n_\B (4 \pi /3) R_\QN^3$, which gives 
\begin{align}\label{eq:R}
	R_\QN \simeq 
%	\bigl( 0.024 \cm \bigr) 
	\bigl( 0.02 \cm \bigr) 
	\left( \frac{N_\B}{10^{34}} \right)^{1/3} \left[ \frac{B}{(150 \MeV)^4} \right]^{-1/4} 
	\per
\end{align}
Comparing with \eref{eq:Ri}, we observe that the radius shrinks by a factor of $\sim 10^3$ as the pocket of unbroken phase cools and reaches to an equilibrium degenerate Fermi state.
The mass of a quark nugget with baryon number $N_\B$ is given by \eref{eq:M_QN}, and we estimate $M_\QN \simeq 9.8 \, B^{1/4} \, N_\B$, which gives
\begin{align}\label{eq:M}
	M_\QN \simeq 
%	\bigl( 1.3 \times 10^{23} \Msun \bigr) 
%	\bigl( 2.6 \times 10^{10} \gram \bigr) 
	\bigl( 3 \times 10^{10} \gram \bigr) 
\left( \frac{N_\B}{10^{34}} \right) \left[ \frac{B}{(150 \MeV)^4} \right]^{1/4} 
	\com
\end{align}
or equivalently $1 \times 10^{-23} \Msun$.  
For reference, note that the Schwartzchild radius for this mass is $R_s = 2 G_N M \simeq 4 \times 10^{-16} \cm$, which makes gravitational effects irrelevant on these scales. 

%=========
The mass and radius of a quark nugget with baryon number $N_\B$ are shown in \fref{fig:mass_radius}.  
Note that an $O(10)$ uncertainty in $R_i$ (discussed above) becomes an $O(10)$ uncertainty in $R_\QN$ and an $O(10^3)$ uncertainty in $M_\QN$.  

%-------------------------------------------
% First order quark-hadron phase transition
%-------------------------------------------
\subsection{First order quark-hadron phase transition}
\label{sec:QCD_Trans}

%=========
The Standard Model predicts that the quark-hadron transition at $\mu_\B = 0$ is a continuous crossover, and lattice simulations infer the approximate temperature to be $T_\QCD \simeq 164 \pm 2 \MeV$ \cite{Fodor:2001pe}.  
However, it is well-known that the quark-hadron transition would be first order if the number of light flavors of quarks were greater than or equal to the number of colors, {\it i.e.} $N_f \geq N_c$ \cite{Pisarski:1983ms}.
If the electroweak phase transition is supercooled below the scale of quark confinement, $\Lambda_\QCD$, then there are effectively $N_f = 6$ light quark species for $N_c = 3$ QCD, and the quark-hadron phase transition is predicted to be first order.\footnote{See also the lattice QCD simulation for the case of $N_f=4$ and $N_c=3$ QCD in \rref{Fodor:2001au}.}  

%=========
To our knowledge there have not been any lattice studies of a quark-hadron phase transition with six flavors of massless quarks.  
Therefore, we will estimate the phase transition temperature, $T_\QCD$, with analytical methods.  
Based on the chiral Lagrangian, the quark condensate $\langle \bar{q} q \rangle_T$ has a temperature dependence given by \cite{Gasser:1986vb} 
\begin{align}\label{eq:qq}
	\langle \bar{q} q \rangle_T = \langle 0 | \bar{q} q | 0 \rangle \biggl[ 1 - \frac{N_f^2-1}{N_f} \Bigl( \frac{T^2}{12 f_\pi^2} \Bigr) - \frac{N_f^2-1}{2N_f^2} \Bigl( \frac{T^2}{12 f_\pi^2} \Bigr)^2 + O(T^6) \biggr] 
	\com
\end{align}
where $\langle 0 | \bar{q} q | 0 \rangle$ is the vacuum quark condensate, and $f_\pi$ is the pion decay constant.  
We define the temperature of the quark-hadron phase transition by $\langle \bar{q} q \rangle_{T_\QCD} = 0$, which lets us estimate 
\begin{align}\label{eq:T_formula}
	T_\QCD 
	\approx \sqrt{ \frac{12N_f}{N_f^2-1} } \biggl( 1 - \frac{1}{4 N_f^2} \biggr) f_\pi 
	\per 
\end{align}
Note that $f_\pi$ will also depend on $N_f$ in general.  

%=========
The result in \eref{eq:T_formula} agrees with the observation made in \rref{Fodor:2001pe} that increasing the number of light quarks lowers the phase transition temperature.  
For $N_f = 3$ and $f_\pi = 93 \MeV$ we obtain $T_\QCD \simeq 190 \MeV$, which is a reasonably good approximation to the lattice result, $T_\QCD \simeq 164 \MeV$ \cite{Fodor:2001pe}, and the two values differ by only $\sim 15\%$.    
Using $N_f = 6$ in \eref{eq:T_formula} gives 
\begin{align}\label{eq:T_QCD}
	T_\QCD \simeq 130 \MeV \, \left( \frac{f_\pi}{93 \MeV} \right)
	\com
\end{align}
and we expect an uncertainty of $O(15\%$) like the $N_f = 3$ case.  
Note that for $N_f = 6$ QCD, the value of $f_\pi$ may differ from its canonical value.  

%=========
Note that \eref{eq:qq} should not be taken literally.  
This formula describes a second order phase transition in which $\langle \bar{q} q \rangle_T$ is continuous across $T_c$ but $(d/dT) \langle \bar{q} q \rangle_T$ is discontinuous.  
We only use \eref{eq:qq} to derive a rough estimate of the phase transition temperature, and not the order of the phase transition.  

%-------------------------------------------
% Supercooled electroweak phase transition
%-------------------------------------------
\subsection{Supercooled electroweak phase transition}
\label{sec:EW_Trans}

%=========
The Standard Model predicts that the electroweak phase transition is a continuous crossover at a temperature of $T_\EW \simeq 160 \GeV$ \cite{DOnofrio:2015mpa}.  
However, it is well-known that the presence of new physics at the weak scale can dramatically change the nature of the electroweak phase transition, possibly lowering the transition temperature or even causing it to become a first order phase transition~\cite{Cohen:1993nk,Carena:1996wj}.  
Here we are interested in new physics that allows $T_\EW < T_\QCD \sim 130 \MeV$ so that the electroweak symmetry is unbroken at the time of the quark-hadron transition.  

%=========
Two groups \cite{Iso:2017uuu,Arunasalam:2017ajm} have recently studied models of the electroweak phase transition with extreme supercooling (see also \rref{vonHarling:2017yew}).  
The basic idea is as follows.  
The Higgs theory is extended to include an additional scalar field such that the potential  has an approximate flat direction.  
Specifically, the flat direction runs through the origin where the leading terms in the potential are quartic.  
When the system is brought to finite temperature, the thermal mass corrections stabilize the potential at the origin.  
Since the thermal mass corrections do not have to compete against a tachyonic mass parameter (as in the Standard Model), the origin remains a local minimum to very low temperatures, and the electroweak phase transition can experience dramatic supercooling.  
Provided that the electroweak phase transition is supercooled to $T_\EW < T_\QCD \sim 130 \MeV$, then the quark-hadron phase transition is expected to be first order, as we discussed in \sref{sec:QCD_Trans}.  

%=========
To be more concrete and provide a proof of principle, we will briefly present a benchmark model \cite{Iso:2017uuu}.  
Let $h = \sqrt{2} | \Phi |$ denote the Higgs field, and let $\phi$ be a new, real scalar field.  
The scalar potential is written as a sum of the tree-level contribution and the one-loop Coleman-Weinberg correction (including the possibility of matter that couples to both the Higgs and the singlet):
\begin{align} \label{eq:CW-potential}
	V(h,\phi) & = V_0 + \frac{1}{4} \lambda_h \, h^4 + \frac{1}{4} \lambda_{\rm mix} \, h^2 \phi^2 + \frac{1}{4} \lambda_\phi \, \phi^4 
	+ \frac{1}{4} B_h \, h^4 \left( \log \frac{h}{v} - \frac{1}{4} \right) 
	\nn & \quad 
	+ \frac{1}{4} B_\phi \, \phi^4 \left( \log \frac{\phi}{M} - \frac{1}{4} \right) 
	+ \frac{1}{4} B_{h\phi} \, \bigl( h^2 + \kappa \, \phi^2 \bigr)^2 \left( \frac{1}{2} \log \frac{h^2 + \kappa \, \phi^2}{v^2 + \kappa \, M^2} - \frac{1}{4} \right) 
	\per
\end{align}
Note that the mass parameters are absent.\footnote{This parameter point is not radiatively stable; the mass parameters are generated by interactions between the Higgs and new, heavy particles.  In other words, a fine-tuning is required to ensure that the mass parameters vanish.  Admittedly, this tuning is no worse than the standard gauge hierarchy (Higgs naturalness) problem, but arguably it is more reasonable to tune the mass corrections to zero rather than the weak scale. See \rref{Bardeen:1995kv} for the possibility of using classical scale invariance to explain the naturalness problem and \rref{Tavares:2013dga} for the counter argument.}  
For $\lambda_{\rm mix} = - 2 \lambda_h v^2 / M^2$ and $\lambda_\phi = \lambda_{\rm mix}^2 / ( 4 \lambda_h )$ the tree-level potential has a flat direction along which $\phi = \sqrt{ -2 \lambda_h / \lambda_{\rm mix}} \, h$.  
The term $V_0 = B_\phi M^4 / 16 + B_h v^4 / 16 + B_{h\phi} \bigl( v^2 + \kappa \, M^2 \bigr)^2 / 16$ ensures that the scalar potential has a global minimum at $(h,\phi) = (v,M)$ where $V(v,M) = 0$.  

%The coupling of the Higgs to the SM particles leads to 
%\begin{align}\label{eq:Bh_SM}
%	B_h \approx \frac{-12 m_t^4 + 6 m_W^4 + 3 m_Z^4 + 9 \lambda_h^2 v^4}{8\pi^2 v^4} \simeq - 0.034 
%	\com 
%\end{align}
%and the Higgs-singlet coupling contributes an additional term of order $B_h \sim \lambda_{\rm mix}^2$, which is negligible for the parameter regime of interest.  

\begin{figure}[t]
\begin{center}
\includegraphics[width=0.99\textwidth]{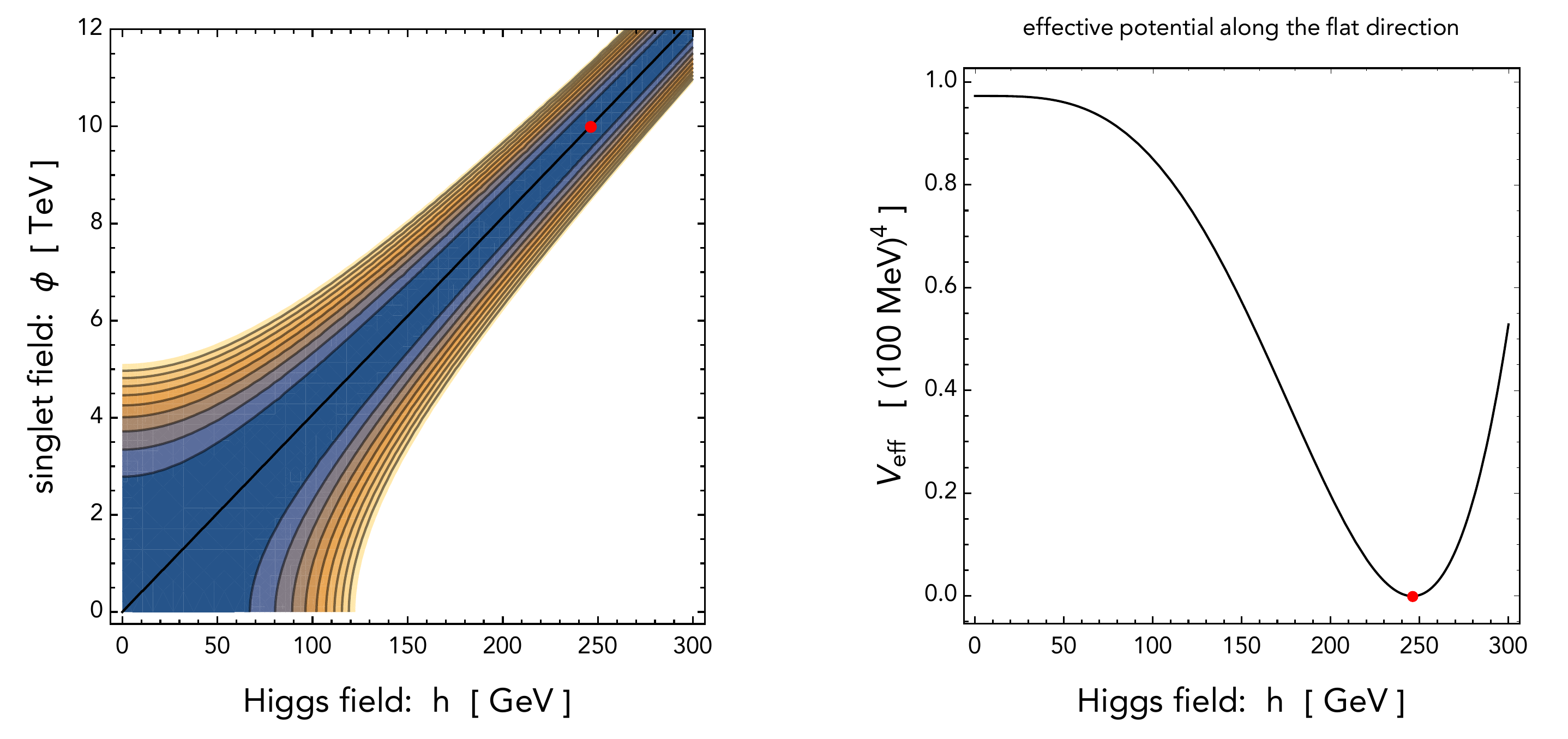}
\caption{
\label{fig:effective_potential} The left panel shows a contour plot of the effective potential as a function of $h$ and $\phi$.  The diagonal direction is an approximately-flat direction. The red point is the location of the global vacuum with $h=246 \GeV$ and $\phi=10 \TeV$.  The right panel shows the effective potential along the flat direction with $\phi = \sqrt{ -2 \lambda_h / \lambda_{\rm mix}} \, h$ as a function of the Higgs field.
}
\end{center}
\end{figure}

%=========
A benchmark model is specified by the parameter choices:
\begin{align}\label{eq:benchmark}
	& \lambda_h = 0.146 
	\, , \ \ 
	\lambda_\phi = 5.34679193376 \times 10^{-8} 
	\, , \ \ 
	\lambda_{\rm mix} = -1.7670672 \times 10^{-4}  \,,
	\nn & 
	M = 10 \TeV 
	\, , \ \ 
	v = 246 \GeV 
	\, , \ \ 
	\kappa = 1 \,,
	\nn & 
	B_h = -0.034 
	\, , \ \ 
	B_\phi = -2.057563413747342 \times 10^{-5} 
	\, , \ \ 
	B_{h\phi} = 2.056319 \times 10^{-5} 
	\per 
\end{align}
The spectrum consists of a heavy scalar boson with mass $m \simeq 125 \GeV$, which is mostly $h$, and a light scalar boson with mass $m \simeq 830 \eV$, which is mostly $\phi$; the mixing is approximately $\tan 2 \theta \simeq -0.05$.  
The vacuum energy is $V_0 \simeq (100 \MeV)^4$.  
The effective potential for this model is shown in \fref{fig:effective_potential}.  

%=========

%=========
As one would expect, this parameter point is highly tuned; all of the digits shown in \eref{eq:benchmark} are needed to obtain the desired solution.  
For instance, if the last few digits of $B_\phi$ or $B_{h\phi}$ are dropped, then $V_0$ grows above $\Lambda_\QCD^4$, and if some digits are removed from $\lambda_\phi$ or $\lambda_{\rm mix}$ then the vacuum shifts away from $(h,\phi) = (v,M)$. 
The parameters are chosen in this way to ensure that $V_0 < T_\QCD^4$, which avoid potential problems with reheating; see the discussion in \sref{sec:overview}.  
Since the ``natural'' value of $V_0$ is $B_\phi M^4 / 16 \sim (300 \GeV)^4$, the required degree of tuning is very high.  
Explanations for such a fine-tuning may rely on the underlying origin of the $\phi$ field, either a dilaton or radion, from some spontaneous symmetry breaking of conformal field theory.

%-------------------------------------------
% Baryon number accumulates in the quark nuggets
%-------------------------------------------
\subsection{Baryon number accumulates in the quark nuggets}
\label{sec:B_number}

%=========
Let us now estimate the relative densities of baryon number in the quark and hadronic phases during the phase transition following \rref{Witten:1984rs}.  
Provided that the wall expands sufficiently slowly, the system can reach thermal and chemical equilibrium at the bubble wall.  
Thermal equilibrium implies that both phases have a common temperature, denoted $T_\QCD$.  
Chemical equilibrium implies that the baryon-number chemical potentials in the hadronic and quark phases are equal, $\mu_\B^\Lph = \mu_\B^\Hph$.  
Just after the phase transition we have $\mu_\B / T_\QCD \ll 1$ in both phases, which lets us write $n_\B = (\mu_\B / T_\QCD) \langle Q_\B^2 \rangle_0$ where $\langle Q_\B^2 \rangle_0$ is the thermal average of the squared baryon-number charge operator at $\mu=0$.  
Therefore the ratio of the baryon number in the two phases is\footnote{In the quark phase, baryon number and lepton number are not separately conserved.  Equating the chemical potentials for the three conserved charges, $\B/3 - \L_i$, gives a ratio $r$ in \eref{eq:r_def} that is smaller by a factor of $5/9$.  We neglect this $O(1)$ numerical factor in the following analysis.  } 
\begin{align}\label{eq:r_def}
	r \equiv \frac{n_\B^\Lph}{n_\B^\Hph} = \frac{\langle Q_\B^2 \rangle_0^\Lph}{\langle Q_\B^2 \rangle_0^\Hph} 
	\per
\end{align}
Next we evaluate $\langle Q_\B^2 \rangle_0$ in the two phases.  

%=========
In the quark phase, $\langle Q_\B^2 \rangle_0^\Hph$ receives contributions from the six flavors of massless quarks, which each have baryon number $1/3$.  
Summing over two spin, two antiparticles, and three colors we find 
\begin{align}\label{eq:QB_H}
	\langle Q_\B^2 \rangle_0^\Hph 
	\approx 6 \times 2 \times 2 \times 3 \times \left( \frac{1}{3} \right)^2 \int \! \! \frac{\ud^3 \pvec}{(2\pi)^3} \frac{1}{e^{|\pvec| / T_\QCD} + 1} 
	= \frac{6 \, \zeta(3)}{\pi^2} T_\QCD^3
	\per
\end{align}
In the hadronic phase it is more subtle to count the degrees of freedom.  
The quark condensate induces a tadpole for the Higgs field that leads to a QCD-scale Higgs condensate, $|\langle \Phi \rangle| \sim v_\QCD \sim 100 \MeV$.  
Since an electroweak-scale Higgs condensate is not present to lift the heavy quark masses, the spectrum contains six flavors of quasi-degenerate, light quarks that confine to form $N_{\rm bary} = 70$ quasi-degenerate baryons ($Q_\B = \pm 1$) with mass $m_B$.  
However eventually the Higgs field reaches its zero-temperature value, $|\langle \Phi \rangle| \sim v_\EW = 246 \GeV$, and there are only $N_{\rm bary} = 2$ light baryons, corresponding to the neutron and proton.  
Rather than studying the dynamical evolution of the Higgs condensate, we will instead treat $N_{\rm bary}$ as a free parameter.  
With this spectrum the squared charge operator is 
\begin{align}\label{eq:QB_L}
	\langle Q_\B^2 \rangle_0^\Lph 
	& \approx N_{\rm bary} \times 2 \times 2 \times  \bigl( 1 \bigr)^2 \int \! \! \frac{\ud^3 \pvec}{(2\pi)^3} \frac{1}{e^{( m_B + |\pvec|^2 / 2m_B ) / T_\QCD} + 1} \nn 
	& \approx 4 N_{\rm bary} \left( \frac{m_B T_\QCD}{2\pi} \right)^{3/2} \, {\rm exp}\bigl[ - m_B / T_\QCD \bigr] 
	\com 
\end{align}
where the approximation is most reliable in the regime $m_B \gg T_\QCD$.  
Using these formulas, the baryon-number ratio is found to be 
\begin{align}\label{eq:r}
	r \approx N_{\rm bary} \, \frac{\sqrt{2\pi}}{6 \, \zeta(3)} \Bigl( \frac{m_B}{T_\QCD} \Bigr)^{3/2} \, {\rm exp}\bigl[ - m_B / T_\QCD \bigr] 
	\com
\end{align}
which only depends on the ratio of baryon mass to phase transition temperature, $m_B / T_\QCD$, and the number of light baryons in the hadronic phase, $N_{\rm bary}$.    

%=========
To evaluate $r$ we must first determine the ratio $m_B / T_\QCD$ for $N_f = 6$ flavors of light quarks.  
The temperature is inferred from \eref{eq:T_QCD}, which gives $T_\QCD \simeq 1.4 f_\pi$ for $N_f = 6$.  
The baryon mass $m_B$ has been measured on the lattice in a theory of QCD with $N_f = 6$ to obtain $m_B / f_\pi \simeq 11 \pm 1$ \cite{Appelquist:2010xv,Appelquist:2014zsa}.  
Using the ratio $m_B / T_\QCD \simeq 11 / 1.4 \simeq 7.8$ and $N_{\rm bary}=70$ we obtain $r \simeq 0.22$.  
For comparison, a model with three flavors of light quarks would have $m_B / T_\QCD \simeq 938 \MeV / 164 \MeV \simeq 5.7$ instead and we obtain $r \simeq 0.03$ with $N_{\rm bary} = 2$. 

%-------------------------------------------
% 6FQM nugget relic abundance
%-------------------------------------------
\subsection{6FQM nugget relic abundance}
\label{sec:Ratio}

%=========
The localized nuggets of quark matter may survive in the universe today and thereby provide a candidate for dark matter.  
In this section we calculate the relic abundance of quark matter assuming that they are cosmologically stable, and we discuss the issue of stability further in \sref{sec:Stability}.  
Since the cosmological production of quark matter requires the baryon asymmetry of the universe to be nonzero at the QCD phase transition, it is therefore natural to ask whether this model of dark matter can also explain the relative abundances of dark matter and baryons, which takes the value $\Omega_\DM / \Omega_b \simeq 5.3$ in the universe today~\cite{Ade:2015xua}.  

%=========
In this section, we calculate the quantity $\Omega_\QN / \Omega_b$.  
Let $\rho_\QN(t)$ be the cosmological energy density of the quark nuggets at time $t$ after the QCD phase transition, and let $\rho_b(t)$ be the energy density of free baryons (not bound in quark matter) at time $t$.  
We are interested in 
\begin{align}
	\frac{\Omega_\QN}{\Omega_b} = \frac{\rho_\QN(t_0)}{\rho_b(t_0)} 
	\com
\end{align}
where $t_0$ is the age of the universe today.  
Both the quark nuggets and the free baryons are non-relativistic today, and we can write 
\begin{align}
	\rho_\QN(t_0) = M_\QN \, n_\QN(t_0)   \com 
	\qquad \text{and} \qquad 
	\rho_b(t_0) = m_p \, n_b(t_0) \com 
\end{align}
where $M_\QN$ is the mass of a quark nugget given by \eref{eq:M_QN}, $n_\QN(t)$ is the cosmological density of quark nuggets at time $t$, $m_p$ is the proton mass, and $n_b(t)$ is the cosmological density of free baryon number at time $t$.  
(As a simplifying approximation we assume that all quark nuggets have the same mass.)
Assuming that the quark nuggets are cosmologically stable and that baryon number is conserved after the QCD epoch, we can write 
\begin{align}
	\frac{n_\QN(t_0)}{n_b(t_0)} = \frac{n_\QN(t_\QCD)}{n_b(t_\QCD)}
	\per
\end{align}
The variable $n_b(t_\QCD)$ is precisely the quantity that we denoted as $n_\B^\Lph(t_\QCD)$ in \sref{sec:B_number}.  
Recall from \eref{eq:r_def} that $n_\B^\Lph = r \, n_\B^\Hph$ where a formula for $r$ appears in \eref{eq:r}.  
The variable $n_\B^\Hph(t_\QCD)$ is just the number density of baryon number inside of a quark nugget at the end of the QCD phase transition.  
We can write $n_\B^\Hph(t_\QCD) = N_\B(t_\QCD) / V_\QN(t_\QCD)$ where $N_\B(t)$ is the total baryon number in the quark nugget at time $t$, and $V_\QN(t)$ the quark nugget's volume.  
To summarize the calculation, we have derived 
\begin{align}
	\frac{\Omega_\QN}{\Omega_b} 
	= \frac{M_\QN}{N_\B(t_\QCD)} \frac{1}{r \, m_p} \Bigl[ n_\QN(t_\QCD) V_\QN(t_\QCD) \Bigr]  \com 
\end{align}
where $M_\QN$ and $r$ are given by \erefs{eq:M_QN}{eq:r}.  The product $n_\QN V_\QN$ measures the number density of quark nuggets in the Hubble volume at the QCD epoch, multiplied by the volume of a given quark nugget; this product can be also interpreted as the fraction of space occupied by the quark nuggets at the end of QCD phase transition. This product is anticipated to be order one and must be $\leq 1$, since quark nuggets do not overlap.
Assuming that $N_\B(t)$ is the same at the end of the QCD phase transition and at the time when equilibrium (pressure balance) is reached, then the factors of $N_\B$ cancel out in the ratio $M_\QN / N_\B(t_\QCD)$, and we finally obtain 
\begin{align}\label{eq:Omega_ratio}
	\frac{\Omega_\QN}{\Omega_b} 
	& \simeq \bigl( 7.2 \bigr) \left( \frac{B^{1/4}}{150 \MeV} \right) 
	\left( \frac{N_{\rm bary}}{70} \right)^{-1} \Bigl( \frac{m_B/T_\QCD}{7.8} \Bigr)^{3/2} \, e^{-(m_B / T_\QCD - 7.8) } \Bigl[ n_\QN(t_\QCD) V_\QN(t_\QCD) \Bigr] 
	\per
\end{align}

%=========
\begin{figure}[t]
\begin{center}
\includegraphics[width=0.65\textwidth]{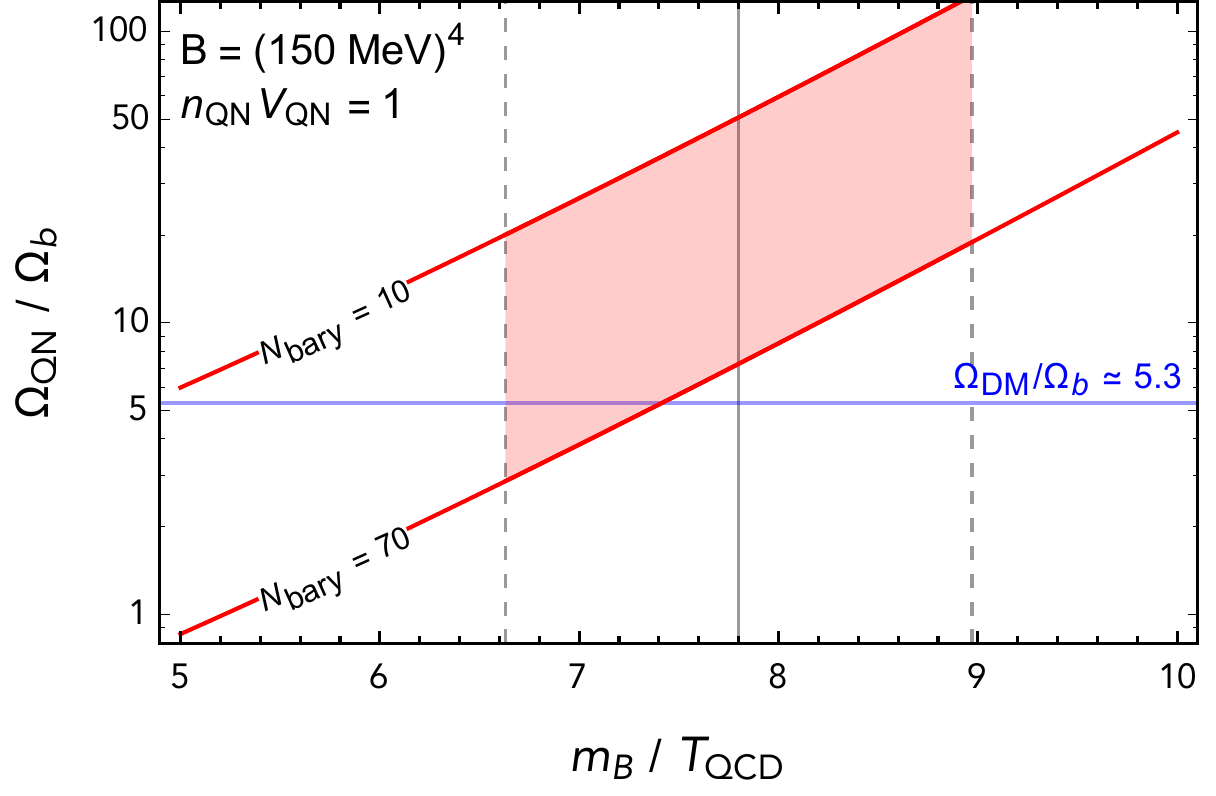} 
\caption{
\label{fig:relic_abund}
The relic abundance of 6FQM nuggets as a function of the baryon mass $m_B$ and the QCD phase transition temperature $T_\QCD$.  In \sref{sec:B_number} we estimate $m_B / T_\QCD \simeq 7.8$, and we show the $\pm 15\%$ error bands.  We fix the differential vacuum pressure to be $B \simeq (150 \MeV)^4$, which matches the usual QCD bag parameter.  The number of light baryons in the hadronic phase at $t_\QCD$ depends on the value of the Higgs condensate at that time; the value of $N_{\rm bary}$ varies from $10$ (top) to $70$ (bottom).  
}
\end{center}
\end{figure}

%=========
The quark nugget-to-baryon ratio is shown in \fref{fig:relic_abund} as a function of the ratio $m_\B / T_\QCD$.  
Recall that $m_B$, the baryon mass in the hadronic phase at $t_\QCD$,  can differ from the usual proton mass scale, because the Higgs condensate may not yet have reached its zero-temperature value.  
At the end of \sref{sec:B_number} we estimated $m_\B / T_\QCD \simeq 7.8$ by adapting results from chiral perturbation theory and lattice studies, but we expect at least an $O(10\%)$ uncertainty in these estimates.  
For the expected number of light baryons, $N_{\rm bary} = 70$, it is remarkable and encouraging that the predicted relic abundance of six-flavor quark matter nuggets falls within an $O(1)$ factor of the measured dark matter relic abundance, $\Omega_\DM / \Omega_b \simeq 5.3$.  
Going forward, more work needs to be done to reduce the large uncertainties in the calculation, particularly to more carefully estimate the values of $N_{\rm bary}$ and $(n_\QN V_\QN)$.  

%==================================
% Stability
%==================================
\section{Stability}
\label{sec:Stability}

%=========
To investigate whether the quark matter phase is stable, we compare its energy against the energy of a system in the hadronic phase with equal values for the conserved charges (electromagnetic charge, $\B$, and $\L$).  
In the hadronic phase at $T=0$, the lowest energy charge-neutral configuration would consist of $n_\B$ protons, $n_\B$ electrons, and $n_\B-n_\L$ antineutrinos at rest.  
The energy density of this configuration is $m_p n_\B + m_e n_\B + m_\nu (n_\B-n_\L)$ where $m_p \simeq 938 \MeV$ is the proton mass, $m_e \simeq 0.511 \MeV$ is the electron mass, and $m_\nu < 1 \eV$ is the neutrino mass scale.  
Since $n_B$ and $n_\L$ are comparable in magnitude [see below \eref{eq:equilibrium}], the energy density is dominated by the first term, $m_p n_\B$.  
Therefore we can define the stability parameter 
\begin{align}\label{eq:S_def}
	\Scal 
	\equiv \frac{\rho}{m_p \, n_\B} 
%	= \left( \frac{71473375 \, \pi^2 B}{76832 \, m_p^4} \right)^{1/4} 
	\simeq 9.8 \frac{B^{1/4}}{m_p}
	\com 
\end{align}
which is the binding energy of six-flavor quark matter per unit baryon number per unit proton mass.  
A value $\Scal < 1$ ensures that the quark matter phase is stable.  

%=========
To evaluate $\Scal$ we need a value for the differential vacuum pressure, $B$.  
In general $B$ receives contributions from both the quark condensate ($\langle \bar{q} q \rangle \neq 0$) and the Higgs condensate ($\langle \Phi \rangle \neq 0$), and we can write $B = B_{\bar{q}q} + B_\Phi$ where both terms are positive.  
We have discussed a model for the electroweak phase transition in \sref{sec:EW_Trans} where we have seen that $B_\Phi < B_{\bar{q}q}$ is possible with sufficient tuning.  
Therefore we neglect $B_\Phi$ and approximate $B \approx B_{\bar{q}q}$.  
For three-flavor QCD the parameter $B_{\bar{q}q} = B_{\bar{q}q,3}$ is identified with the bag parameter of the MIT bag model, and its value is determined from measurements of the spectrum of light hadrons to be $B_{\bar{q}q,3} \simeq (150 \MeV)^4$ \cite{Hasenfratz:1977dt}.  
Although it is not clear that $B_{\bar{q}q,3}$ will be the same for six-flavor quark matter, this value represents a useful benchmark, which gives $\Scal = 1.57 (B^{1/4} / 150 \MeV)$.  
The value $\Scal > 1$ implies that the six-flavor quark matter is unstable; the system could lower its energy by fragmenting into free nuclei and leptons.  
For comparison, one obtains $\Scal \simeq 0.91 (B^{1/4} / 150 \MeV)$ for three-flavor quark matter where $\rho = 4B$ and $n_\B \simeq 0.7 B^{3/4}$ \cite{Farhi:1984qu}.  

%-------------------------------------------
\subsection{Six-flavor quark matter is metastable: lifetime estimate}

%=========
Although $\Scal > 1$ for six-flavor quark matter with $B = (150 \MeV)^4$, we now argue that the quark matter is metastable and long lived.  
As with the $\alpha$-decay of some heavy nuclei, we expect that the quark matter decay will be mediated by a (thermal) ``tunneling'' process with a suppressed rate.  
Instead of the Coulomb barrier for $\alpha$-decays, the phase boundary could also provide a potential barrier to reduce the emission rate of a baryon from the quark matter.  
For the $\alpha$-decay case, the $\alpha$ particle separation distance from the bulk nucleus is used to derive an effective potential.  
Similarly, we will use the energy of the quark matter as a function of the radius to model the effective potential.  

%=========
The energy density of the quark matter is given by \eref{eq:rho_def}, which can be written as $\rho = (1245/8\pi^2) \mu^4 + B$, and the chemical potential is given by $\mu = (\pi^2 n_\B / 14)^{1/3}$ from \eref{eq:n_sol}.  
Now consider a spherical region of quark matter with radius $R$ and total baryon number $N_\B$.  
The energy of this system is 
\begin{align}\label{eq:E_of_R}
	E(R) = \rho V = \biggl[ \frac{1245}{8\pi^2} \left( \frac{\pi^2}{14} \right)^{4/3} \left( \frac{4}{3} \pi \right)^{-1/3} N_\B^{4/3} \biggr] \frac{1}{R} + \biggl[ \frac{4}{3} \pi B \biggr] R^3 
	\per
\end{align}
For fixed values of $N_\B$ and $B$, the energy has a minimum at
\begin{align}\label{eq:Emin_Rmin}
%	E_{\rm min} = \frac{415^{3/4} \pi^{1/2}}{14 \times 2^{1/4}} N_\B B^{1/4}  \,,
	E_{\rm min} \simeq 9.8 N_\B B^{1/4} 
	\qquad \text{and} \qquad
%	R_{\rm min} = \frac{3^{1/3} \, 415^{1/4}}{7^{1/3} 2^{7/4} \pi^{1/6}} N_\B^{1/3} B^{-1/4} 
	R_{\rm min} \simeq 0.84 N_\B^{1/3} B^{-1/4} 
	\per
\end{align}
Note that $E_{\rm min} / (4\pi R_{\rm min}^3/3) = 4B$ in agreement with \eref{eq:equilibrium}.  

%=========
\begin{figure}[t]
\begin{center}
\includegraphics[width=0.65\textwidth]{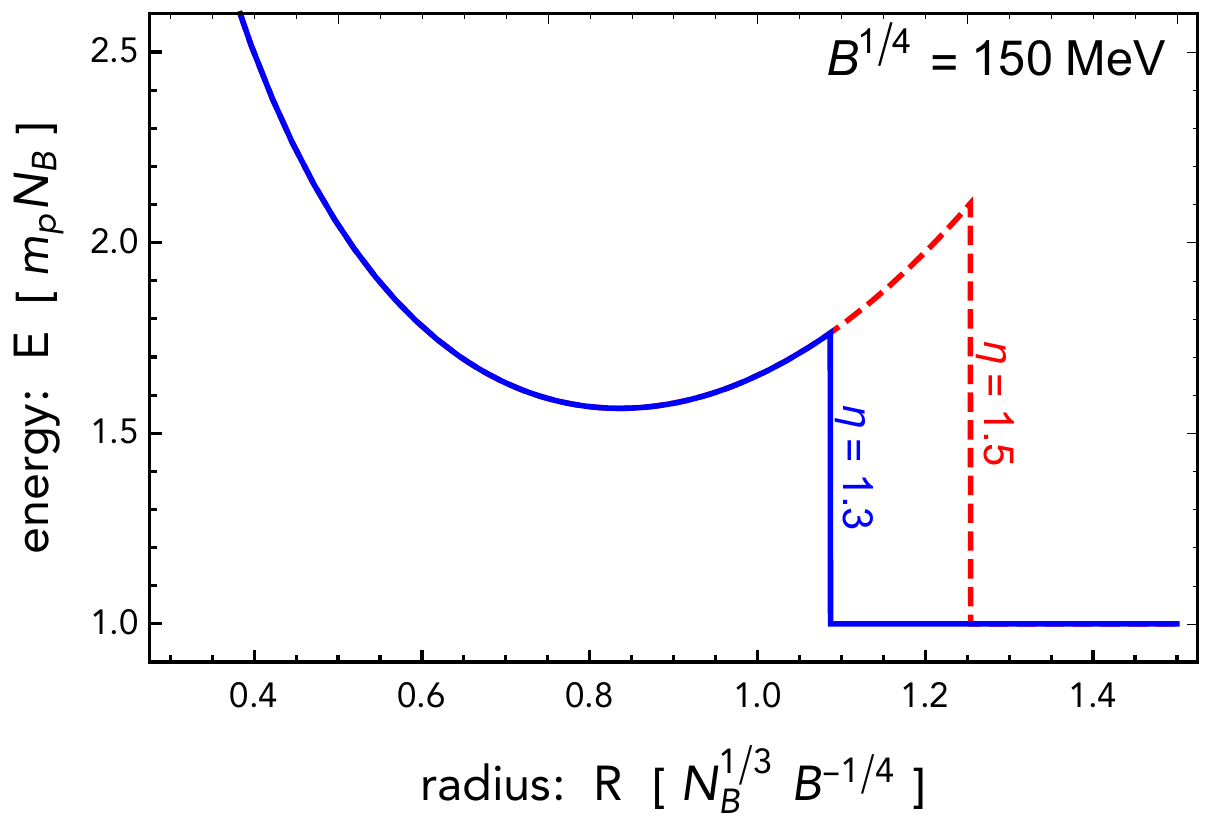}
\caption{
\label{fig:E_of_R}
A spherical region of quark matter with radius $R$, baryon number $N_\B$, and differential pressure $B$ has an energy $E$ given by \eref{eq:E_of_R}.  The local equilibrium at $R_{\rm min} \simeq 0.84 N_{\B-\L}^{1/3} B^{-1/4}$ is unstable toward a tunneling into free hadronic matter where $E \approx m_p N_\B$.  The barrier height depends on the critical radius $R_c = \eta R_{\rm min}$ where the calculation breaks down.  
}
\end{center}
\end{figure}

%=========
When we interpret the energy of the quark matter region as a function of $R$ we find that it has a single, local minimum at $R = R_{\rm min}$ as shown in \fref{fig:E_of_R}.  
For larger values of $R$ the energy grows like $E(R) \sim R^3$, but we expect that this calculation breaks down for $R > R_c$ where $\mu < \mu_c$, and it becomes energetically preferable for the system to pass into the hadronic phase.  
From dimensional analysis we expect
\begin{align}\label{eq:Rc}
	R_c = \eta \, R_{\rm min} 
	\com
\end{align}
where $\eta > 1$ is an order one number.  
The value of $\eta$ controls the height of a potential energy barrier that separates the quark matter phase at $R = R_{\rm min}$ from the free hadronic phase at $R > R_c$.  
The barrier height is expressed as $\Delta E \equiv [E(R_c) - E(R_{\rm min})]/N_\B \simeq 2.4 (3/\eta - 4 + \eta^3)  B^{1/4}$.  

%=========
For the 3FQM, a region of quark matter with baryon number $N_\B + 1$ can have a neutrino-induced decay into quark matter with baryon number $N_\B$ plus a free neutron.  For the 6FQM, the quark nugget is a meta-stable state and has an energy per baryon larger than a free nucleon. So, it can emit baryons just through thermal activation.  
The presence of a potential energy barrier at $R_c$ implies that an energy injection of $\Delta E$ is needed in order to kick quarks out of the quark matter to form a free hadron.  Following \rref{Alcock:1985vc}, the rate is estimated as 
\begin{align}\label{eq:Gamma}
	\Gamma[(N_\B + 1) \rightarrow N_\B + n] \approx \frac{1}{2\pi^2} m_n \, T^2 \,e^{-  \Delta E / T} \,f_n \, \sigma_0 \, N_\B^{2/3}   \com 
\end{align}
where $m_n \simeq 940 \MeV$ is the neutron mass, $f_n \lesssim 1$ is the neutron absorption efficiency, and the geometric cross section of the quark matter is $\sigma = 4 \pi R_{\rm min}^2$, which we write as $\sigma = \sigma_0 N_\B^{2/3}$, which defines $\sigma_0$, and \eref{eq:Emin_Rmin} gives $\sigma_0 \simeq ( 4 \times 10^{-4} \MeV^{-2}) [B / (150 \MeV)^4]^{-1/2}$.  

%=========
The quark matter is metastable provided that its lifetime is longer than the age of the universe.  
The neutron emission rate in \eref{eq:Gamma} causes $N_\B$ to decrease according to $dN_\B / dt = - \Gamma$.  
Therefore a region of quark matter with initial baryon number $N_\B$ has a lifetime
\begin{align}\label{eq:tau}
	\tau(N_\B) = \frac{6 \pi^2}{m_n \, T^2 \,e^{-  \Delta E / T} \,f_n \, \sigma_0} N_\B^{1/3} 
	\per
\end{align}
In the radiation dominated era, the age of the universe is $t = 1/(2H)$ where $H = \sqrt{\pi^2 g_\ast / 90} \ T^2 / \Mpl$ is the Hubble expansion rate, $\Mpl \simeq 2.43 \times 10^{18} \GeV$ is the reduced Planck mass, and $g_\ast \simeq 20$ is the approximate effective number of relativistic species just after the QCD transition.  
Requiring $\tau > t$ gives a lower bound on the baryon number 
\begin{align}\label{eq:NB}
	N_\B 
	 & >  \frac{5^3\,3^2\,83^{3/2}}{2^9\,7^2\,\pi^7} \, \frac{f_n^3 m_n^3 \Mpl^3}{g_{\ast}^{3/2} B^{3/2} } \ {\rm exp}\bigl[ - 3 \Delta E / T \bigr]   \nn
	& \simeq \bigl( 1 \times 10^{30} \bigr) \left( \frac{f_n}{1} \right)^3 \left( \frac{g_\ast}{20} \right)^{-3/2} \left[ \frac{B}{(150 \MeV)^4} \right]^{-3/2} \, {\rm exp}\left[ - 3 \bigl( \Delta E / T - 20 \bigr) \right] 
	\per
\end{align}
\eref{eq:NB} reveals that a region of quark matter with sufficiently large $N_\B$ can be metastable. 

%=========
In the above estimates, the possible important reabsorption of emitted hadrons has been ignored.  
This effect depends on the effective binding energy of hadrons in a thin surface layer and can make the lifetime of the quark matter a few orders of magnitude longer \cite{Madsen:1986jg}. 
Finally, there is also another possible evaporation of quark matter via the ``boiling" effect, for which hadronic gas can be formed inside the bulk of the quark matter.  
Depending on the surface tension of the quark matter and the size of the hadronic gas bubbles, those bubbles could grow and convert the quark matter into nucleons~\cite{Alcock:1988br}.  
For three-flavor quark matter, the boiling effect was studied in more detail in Refs.~\cite{Olesen:1991zt,Olesen:1993iz} using Walecka's mean-field theory for an interacting hadronic gas, and it was argued that the boiling effect is not sufficient to evaporate the quark matter.  
While we expect these arguments to carry over for six-flavor quark matter, it would be useful to revisit this work with a more careful calculation of the barrier height and thermal activation rate for the decay of 6FQM phase into hadronic phase.  

%-------------------------------------------
\subsection{Six-flavor quark matter survives at $T = T_\QCD$}
\label{sec:stability_highT}

%=========
The previous stability argument assumes that $T \ll  \Delta E$ in the hadronic phase outside of the quark matter.  
One may worry that the quark matter is destabilized already at temperatures $T_\QCD \sim \Lambda_\QCD$ just after the QCD phase transition, and we now argue that this is not the case.  

%=========
During the QCD phase transition, the quark condensate develops in the hadronic phase, which induces a vacuum expectation value for the Higgs field $|\langle \Phi \rangle| = v_\QCD \sim \Lambda_\QCD$.  
The expectation value does not grow out to $v_\EW \simeq 246 \GeV$ until the temperature has decreased further.  
Therefore, all of the quark species are light in the hadronic phase at $T_\QCD \sim \Lambda_\QCD$, and the hadron mass spectrum is altered accordingly.  
To know whether the quark matter is in the lowest energy per baryon state or not, we need to know the vacuum pressure $B$ and baryon mass $m_B$ for six-flavor QCD.  

%=========
The vacuum pressure is related to the QCD vacuum energy, which is given by
\begin{align}\label{eq:B_def}
	B \equiv - \langle \Theta^\mu_\mu \rangle \approx - \left\langle \frac{\beta(\alpha)}{4\,\alpha} G^a_{\mu\nu} G^{a\mu\nu} \right\rangle 
	\com 
\end{align}
where the gluon condensation only contains the true non-perturbative contribution after subtracting the perturbative contribution.  
(Here, we have ignored the quark-mass operator contribution to the vacuum energy, which is negligible compared to the gluon condensation.)  
To calculate the precise value of $B$, one need to rely on a non-perturbative method such as lattice QCD.  
For the chiral symmetry breaking, the quark-anti-quark condensation is related to the pseudo-Nambu-Goldstone Boson, pion, mass as $m^2_\pi \,f^2_\pi \simeq 2 m_q \langle \bar{q} q \rangle$ via the partially conserved axial current formula.  
Similarly for the gluon condensation and if the conformal symmetry is a good symmetry, a light dilaton, $0^{++}$, may also exist with its mass related to the condensation via the partially conserved dilatation current  formula, $m_\sigma^2 f^2_\sigma \simeq - 4\,\langle \Theta^\mu_\mu \rangle = 4 B$~\cite{Collins:1976yq,Nielsen:1977sy,Appelquist:2010gy}.  
Here, $m_\sigma$ and $f_\sigma$ are the mass and decay constant of the potential dilaton state, respectively.  
Therefore, we can re-express the ratio of the vacuum energy scale over the baryon mass as 
\begin{align}
	\frac{B^{1/4}}{m_B} \simeq \frac{m_\sigma^{1/2} \, f_\sigma^{1/2} }{2^{1/2}\,m_B} 
	\per
\end{align}

%=========
For $N_f=3$ of the ordinary QCD vacuum case, there is no clear light $0^{++}$ mode.
[The $f_0(500\,\mbox{MeV})$ state has a very broad width, comparable to its mass.]  
As the number of massless flavors increases, the $\SU{3}$ gauge theory becomes more and more conformal in the infrared scale.  
At the transition of the critical number of flavors $N_f^c$, a light dilaton much below the confinement scale is likely to exist.  
For $N_f=8$, lattice simulations from two groups~\cite{Appelquist:2014zsa,Appelquist:2016viq,Aoki:2014oha,Aoki:2016wnc} have shown clear evidence for a light $0^{++}$ with a mass comparable to the pion.  
The current simulation results have a large error on the dilaton mass for the chiral limit of $m_\pi \rightarrow 0$.  
Even though we do not have a precise calculation for the dilaton mass in the chiral limit for $N_f=8$ and $N_f=6$, the parametrical dependence of the ratio $B^{1/4}/m_B$ as a function of $N_f$ should scale as
\begin{align}
	\frac{B_{N_f}^{1/4}}{m_{B_{N_f}}} \propto \frac{N_f^c - N_f}{N_f^c} \com 
\end{align}
when $N_f$ is close to $N_f^c$, {\it i.e.} the conformal window.  
The critical number of flavors, $N_f^c$, is close to but slightly above 8~\cite{DeGrand:2015zxa}.  
Choosing $N_f^c \approx 8$, we have an estimation of the bag parameter for the six-flavor case as
\begin{align}
	\frac{B_6^{1/4}}{m_{B_6}}  \sim0.4 \times \frac{B_3^{1/4}}{m_{B_3}} \approx 0.064 
	\per
\end{align}
Substituting the above ratio into \eref{eq:S_def}, we estimate the stability parameter as 
\begin{align}
	\Scal = \frac{E^{\rm six}/N_B}{m_{B_6}} \sim 0.6 
	\per
\end{align}
A value $\Scal < 1$ means that at high temperature before the Higgs field rolls to its $246 \GeV$ vacuum, the quark matter is likely to have a lower energy than a free baryon in the hadronic phase.  

%==================================
% Phenomenology and Signatures
%==================================
\section{Phenomenology and Signatures}
\label{sec:Phenomenology}

%=========
In this section we suppose that dark matter is composed of 6FQM nuggets, we highlight various aspects of the phenomenology, and we discuss possible signatures.  

%-------------------------------------------
% Stochastic gravitational wave background
%-------------------------------------------
\subsection{Stochastic gravitational wave background}
\label{sec:Grav_Waves}

%=========
A first order cosmological phase transition produces gravitational waves from the collisions of bubbles and the interactions of bubbles with the cosmological medium.  
In this section we discuss the gravitational wave signal that arises from a first order QCD phase transition in the scenario that we have described above \cite{Witten:1984rs}.  

%=========
The gravitational wave spectrum peaks at a frequency $f_{\rm gw}(t)$ at time $t$, and it falls off like a power law at higher and lower frequencies.  
The peak frequency is controlled by the size of the bubbles when they collide, and we recall from \eref{eq:Ri} that the initial bubble radius is estimated as $R_i = O(10 \cm)$.  
Let $\lambda_{\rm gw}(t_\QCD)$ be the length scale of the gravitational waves at the time of the QCD phase transition, and we estimate $\lambda_{\rm gw}(t_\QCD) \sim R_i$.  
Assuming that the universe expands and cools adiabatically, the gravitational wave frequency redshifts as $f_{\rm gw}(t_0) = f_{\rm gw}(t_\QCD) [ a(t_0) / a(t_\QCD) ]^{-1}$ with $a(t_0) / a(t_\QCD) = (T_\QCD / T_0) \, [ g_{\ast S}(t_\QCD) / g_{\ast S}(t_0) ]^{1/3}$ and $T_0 \simeq 0.234 \meV$ and $g_{\ast S}(t_0) \simeq 3.91$.  
The frequency of these gravitational waves today is 
\begin{align}\label{eq:fgw0}
	f_{\rm gw}(t_0) \simeq \bigl( 2.0 \times 10^{-2} \Hz \bigr) \left( \frac{T_\QCD}{130 \MeV} \right)^{-1} \left[ \frac{g_{\ast S}(t_\QCD)}{20} \right]^{-1/3} 
	\left( \frac{R_i}{10 \cm} \right)^{-1} 
	\per
\end{align}
Here $g_{\ast}(t_\QCD)$ represents the number of effective, relativistic degrees of freedom in the plasma, just after the QCD phase transition is completed.  
This estimate implies that the gravitational wave spectrum will peak at a frequency where the LISA space-based gravitational wave interferometer experiment is sensitive: $f \sim 10^{-5} - 10^{-1} \Hz$ \cite{Caprini:2015zlo}.

%=========
Unlike the peak frequency, the {\it amplitude} of the gravitational wave spectrum depends sensitively on the latent heat of the phase transition and the interaction of the bubble with the plasma, which together determine the efficiency of converting the wall's kinetic energy into plasma kinetic energy and gravitational waves \cite{Caprini:2015zlo}.  
The latent heat is parametrized by $\alpha \equiv B / \rho_{\rm rad}$ where $B$ is the differential vacuum energy and $\rho_{\rm rad}$ is the radiation energy density at the phase transition.  
It is customary to distinguish two regimes.  
If the latent heat of the phase transition is large compared to the plasma energy density, then the bubble may enter the {\it runaway regime} where it expands as if it is in vacuum and the wall velocity accelerates toward the speed of light.  
Alternatively, if the latent heat is small, then the bubble is in the {\it non-runaway regime} where an effective friction slows the motion of the wall, which reaches a (possibly relativistic) terminal velocity.  
For the scenario under study in this article, we expect that the first order phase transition will be in the non-runaway regime.  
This is because we have required the vacuum energy to be small to avoid problems with reheating (see the discussion in \sref{sec:overview}), and because the bubble walls interact strongly with the plasma.  
Since earlier studies of gravitational waves from a first order phase transition at the QCD epoch have assumed a runaway \cite{Caprini:2010xv,Schettler:2010dp,Iso:2017uuu,Aoki:2017aws,Chen:2017cyc,Ahmadvand:2017xrw,Ahmadvand:2017tue}, those results are not directly applicable to our model.  

%=========
Therefore we would argue that a more careful calculation of the gravitational wave signal from a first order, non-runaway QCD phase transition is warranted.  
Since the fluid motions are expected to play an important role in the dynamics and gravitational wave generation, a hydrodynamic lattice simulation --- such as the one presented in \rref{Hindmarsh:2015qta} --- may be suitable.  

%-------------------------------------------
% Gravitational lensing and seismic data
%-------------------------------------------
\subsection{Gravitational lensing and seismic data}
\label{sec:Lensing}

\begin{figure}[t]
\begin{center}
%\hspace*{-0.75cm}
\includegraphics[width=0.7\textwidth]{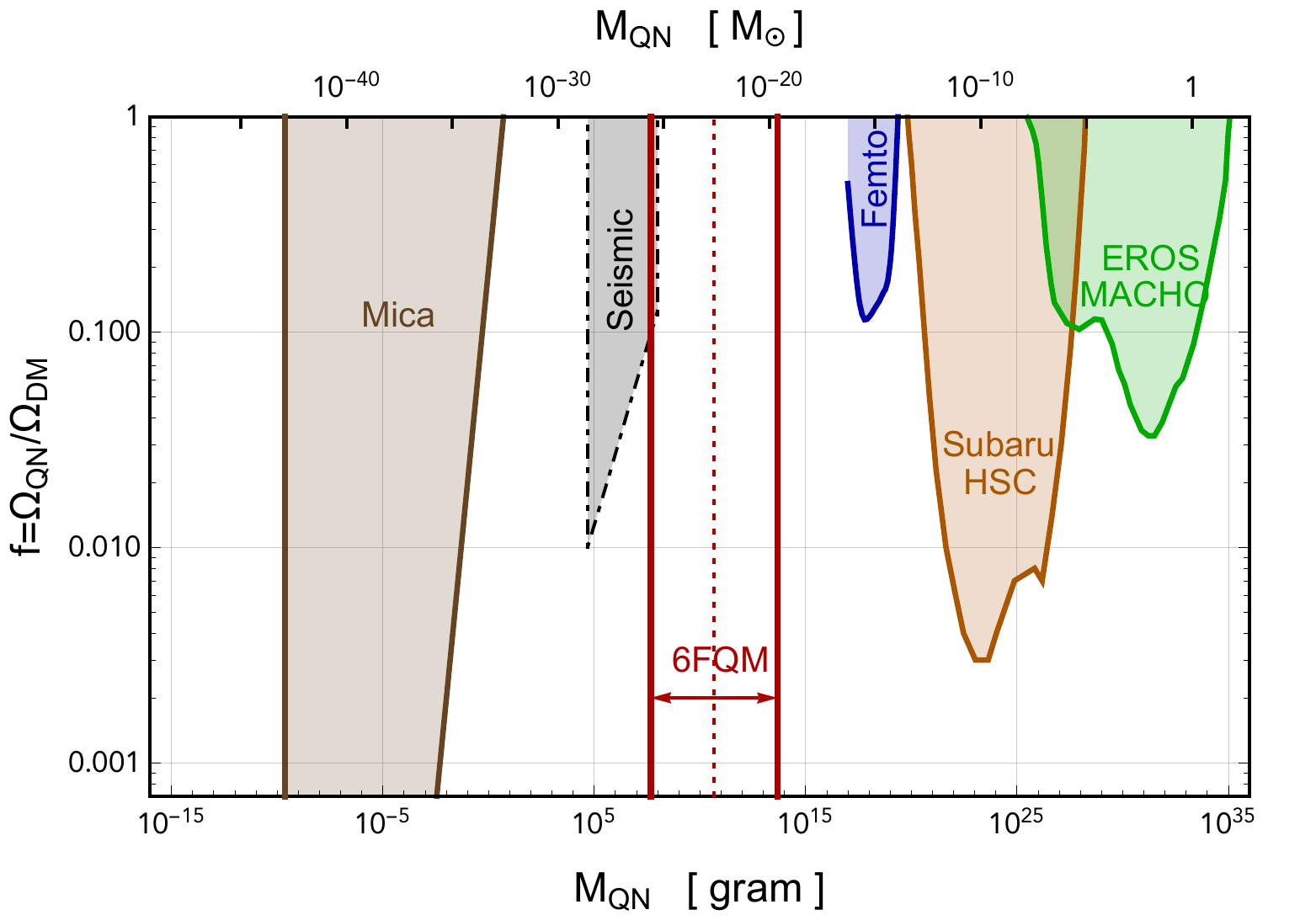}
\caption{The preferred mass range of 6FQM together the existing searches for MACHO from microlensing of EROS and MACHO~\cite{Tisserand:2006zx} and Subaru/HSC~\cite{Niikura:2017zjd} and Femtolensing using Fermi Gamma-ray Burst Monitor (GBM)~\cite{Barnacka:2012bm}. Also shown are the constraint from Mica~\cite{DeRujula:1984axn} and the tentative limits from moon seismic data~\cite{Herrin:2005kb,Cyncynates:2016rij}.  }
\label{fig:macho}
\end{center}
\end{figure}

%=========
The standard way to search for MACHO dark matter is with gravitational lensing.  
For the planetary size objects, the EROS and MACHO collaborations have excluded the possibility of MACHOs as making up all of the dark matter for a wide range of masses from $0.6\times 10^{-7} \Msun$ to $15 \Msun$~\cite{Tisserand:2006zx}.  
For lighter MACHO masses from $10^{-13} - 10^{-6} \Msun$, the Subaru Hyper Suprime-Cam (Subaru/HSC) has set stringent constraints by a 7 hour-long observation of the Andromeda galaxy~\cite{Niikura:2017zjd}.  
For even lighter masses, femtolensing~\cite{1992ApJ...386L...5G} constrains the mass range from $10^{-16}$ to $10^{-13} \Msun$~\cite{Barnacka:2012bm} with measurements of gamma-ray burst energy spectra.  
We summarize the existing experimental constraints in \fref{fig:macho}, and we also show the preferred mass range for 6FQM based on the estimates in \eref{eq:M}.  
The predicted 6FQM mass range falls a few orders of magnitude below the smallest masses that are currently probed by lensing measurements.  

%=========

%=========
The 6FQM mass window presents a new target for gravitational lensing probes of macroscopic dark matter candidates.  
The existing femtolensing analysis~\cite{Barnacka:2012bm} used data from the Fermi Gamma-ray Burst Monitor (Fermi-GBM) to look for the effect of lensing on the spectra of gamma ray bursts at energies $E_\gamma = O(100 \keV)$.  
The Fermi Large Area Telescope (Fermi-LAT) instrument has observed more gamma-ray bursts with good statistics at higher energies, $E_\gamma = O(10 \MeV)$~\cite{Ackermann:2013zfa,vonKienlin:2014nza}.  
Since the femtolensing interference effect roughly needs $E_\gamma\,G_N M \sim 1$, using higher-energy gamma-rays can extend the sensitivity to smaller masses.  
It would be interesting to adapt the analysis of \rref{Barnacka:2012bm} for gamma ray bursts observed by Fermi-LAT; such a study could test the macroscopic dark matter candidates that we have proposed here.  

%=========
Lower-mass dark matter candidates are out of the reach of gravitational lensing measurements, but they can perhaps be probed with seismic data.  
The gray region of \fref{fig:macho} shows possible constraints from Earth and moon seismic data~\cite{DeRujula:1984axn,Herrin:2005kb} in the mass range $10^{-29} - 10^{-26} \Msun$.  
As the quark nugget hits the Earth or moon, it may generate a distinctive linear morphology of seismic events.  
Noticing the mismatch of the small-MACHO size and the long and unattenuated wave-length modes, the updated analysis in \rref{Cyncynates:2016rij} has found no limits for the MACHO with a nuclear density.  
So, the gray region in \fref{fig:macho} should be taken as a possible-future limit from seismic data. 
In the dark brown region for the mass region from $10^{-43}$ to $10^{-33} \Msun$, the search for magnetic monopoles by examining ancient mica for etchable trails of lattice defects~\cite{Price:1983ax} has also been used to place constraints on quark nuggets~\cite{DeRujula:1984axn} (see Ref.~\cite{Jacobs:2014yca} for a recent review about MACHO searches). 

%-------------------------------------------
% Capture by Compact Stars
%-------------------------------------------
\subsection{Capture by compact stars and self-collision}
\label{sec:Capture_Star}

%=========
If a quark nugget is incident upon a star, such as a neutron star or a white dwarf, the quark nugget will very likely be captured.  
In this section we estimate the capture rate, and we discuss the corresponding signal. 
Similarly, we consider the collision of two quark nuggets with each other, and we estimate the possible radiation.  

%=========
Consider a star with mass $M_s$ and radius $R_s$.  
The typical values for neutron stars are $M_s \sim (2-5) \Msun$ and $R_s \sim (5-15) \km$, and for a white dwarf star these values are closer to $M_s \sim (0.2-1.4) \Msun$ and $R_s \sim (0.005-0.02) \Rsun$ \cite{Camenzind:2007}.  
The effective gravitational cross section area is $A_{\rm eff} = (1 + \Theta) A_{\rm geo}$ where $A_{\rm geo} = \pi R_s^2$ is the geometrical cross section area.  
The gravitational enhancement factor, called the Safronov number, is given by $\Theta = v_{\rm esc}^2 / v_{\rm rel}^2$ where $v_{\rm esc} = \sqrt{2 G_N M_s / R_s}$ is the escape velocity at the surface of the star, and $v_{\rm rel}$ is the relative velocity of the dark matter and the star.  
The Safronov number can be as large as $\Theta \sim 10^{6}$ for a neutron star and as large as $10^{2}$ for a white dwarf.  
We assume that every quark nugget that falls within the effective gravitational area is captured.  
The flux of dark matter is given by $\Fcal_{\rm dm} = n_{\rm dm} v_{\rm rel}$ where $n_{\rm dm}$ is the number density of dark matter in the Milky Way halo.  
Since quark nugger dark matter is non-relativistic we can write $n_{\rm dm} = \rho_{\rm dm} / (M_{\QN} c^2)$ where $\rho_{\rm dm} \simeq 0.4 \GeV/{\rm cm}^3$ is the approximate energy density of dark matter in the halo, and $M_{\QN}$ is the mass of a quark nugget.  
Let $N_s$ denote the number of compact stars in the Milky Way; rough estimates for neutron stars and white dwarf stars are $N_s \sim 10^9$ and $N_s \sim 10^{11}$, respectively~\cite{Camenzind:2007}.  
To consider the self-collision of quark nuggets, we estimate the number of nuggets in the Milky Way to be $N_\QN = M_\MW / M_\QN \sim 10^{35}$ for $M_\MW \sim 10^{12} \Msun$ and $M_\QN \sim 10^{10} \gram$.  
Then the average rate at which quark nuggets are captured by stars in the Milky Way or collide with one another is estimated as $\Gamma_{\rm cap} \approx N \Fcal_{\rm dm} A_{\rm eff}$ where $N = N_s$ for stars and $N_\QN$ for nuggets.  
Using the fiducial parameters for neutron stars, white dwarf stars and quark nuggets, we estimate the capture rates as
\begin{subequations}\label{eq:Gam_cap}
\begin{align}
%--
	\Gamma_{\rm cap} \bigr|_{\rm ns}
	& \simeq \bigl( 2 \times 10^{8} \yr^{-1} \bigr) 
	\left( \frac{N_s}{10^9} \right) 
	\left( \frac{R_s}{10 \km} \right) 
	\left( \frac{M_s}{3 \Msun} \right) 
	\left( \frac{M_\QN}{10^{10} \gram} \right)^{-1} 
	\com  \\
%--
	\Gamma_{\rm cap} \bigr|_{\rm wd}
	& \simeq \bigl( 2 \times 10^{12} \yr^{-1} \bigr) 
	\left( \frac{N_s}{10^{11}} \right) 
	\left( \frac{R_s}{0.01 \Rsun} \right) 
	\left( \frac{M_s}{0.5 \Msun} \right)
	\left( \frac{M_\QN}{10^{10} \gram} \right)^{-1}  
         \com  \\
	\Gamma_{\rm cap} \bigr|_\QN
	& \simeq \bigl( 2 \times 10^{13} \yr^{-1} \bigr) 
	\left( \frac{R_\QN}{0.02 \cm} \right)^2 
	\left( \frac{M_\QN}{10^{10} \gram} \right)^{-2}  
	\com
\end{align}
\end{subequations}
where we have taken $v_{\rm rel} = v_{\rm vir} \simeq 300 \km / {\rm sec}$.  
Note that $10^{10} \gram \approx 5.0 \times 10^{-24} \Msun$.  On average, the separation distance from one capture event to another is around 10 pc for neutron stars and 1 pc for white dwarfs and quark nuggets in our galaxy. 

%=========
When the quark nugget reaches the surface of the star, it acquires a kinetic energy $E_{\rm kin} \approx M_\QN v_{\rm esc}^2 /2$ where $v_{\rm esc}$ is the escape velocity at the surface of the star.  
If this energy is liberated as electromagnetic radiation over at time interval $\Delta t$, then it corresponds to a power output $P = f_{\rm rad}\,E_{\rm kin} / \Delta t$ with $f_{\rm rad}$ as the fraction of kinetic energy into radiation energy.  
For the fiducial neutron star and white dwarf parameters, this evaluates to approximately
\begin{subequations}\label{eq:P}
\begin{align}
	P \bigr|_{\rm ns} 
	& \simeq \bigl( 1 \times 10^{-6} \Lsun \bigr) 
	\left(\frac{f_{\rm rad}}{10^{-2}} \right) 
	\left( \frac{M_\QN}{10^{10} \gram} \right) 
	\left( \frac{R_s}{10 \km} \right)^{-1} 
	\left( \frac{M_s}{3 \Msun} \right) 
	\left( \frac{\Delta t}{10 \sec} \right)^{-1} 
	\com \\ 
%--
	P \bigr|_{\rm wd} 
	& \simeq \bigl( 2 \times 10^{-10} \Lsun \bigr) 
	\left(\frac{f_{\rm rad}}{10^{-2}} \right) 
	\left( \frac{M_\QN}{10^{10} \gram} \right) 
	\left( \frac{R_s}{0.01 \Rsun} \right)^{-1} 
	\left( \frac{M_s}{0.5 \Msun} \right) 
	\left( \frac{\Delta t}{10 \sec} \right)^{-1} 
	\com  \\
%--
        P \bigr|_\QN 
        & \simeq \bigl( 1 \times 10^{-12}  \Lsun \bigr) 
        \left(\frac{f_{\rm rad}}{10^{-2}} \right) 
        	\left( \frac{R_\QN}{0.02 \cm} \right)^{-1} 
	\left( \frac{M_\QN}{10^{10} \gram} \right)^{2}  
	\left( \frac{\Delta t}{10 \sec} \right)^{-1} 
	\per 
\end{align}
\end{subequations}
where we have used $\Lsun \simeq 3.83 \times 10^{26} \Watt$.  
Here, we have chosen $f_{\rm rad} = 10^{-2}$, as the situation for a binary neutron star merger~\cite{Hotokezaka:2012ze}. 

%=========
If the quark nugget collides with a white dwarf star, its velocity remains non-relativistic since $\beta = v_{\rm esc}/c \sim 10^{-2}$.  
The deposited kinetic energy will be absorbed by the white dwarf and eventually re-emitted as blackbody radiation.  
Since a white dwarf's average power output is around $P_{\rm wd} \sim 10^{-5} - 10^{-2} \Lsun$, the additional emission that we estimate in \eref{eq:P} is much smaller, and we conclude that QN-WD collisions would be challenging to observe.  
On the other hand, a quark nugget would collide with a neutron star at nearly the speed of light, and the observational signature may be similar to the kilonova that results from a binary neutron star merger, although much dimmer.  
Using GW170817 and GRB 170817A~\cite{TheLIGOScientific:2017qsa,GBM:2017lvd,Monitor:2017mdv} as a reference example, we compare the energy deposited in a QN-NS collision with the radiation energy from a binary neutron star merger, finding that the former is smaller by a factor of $10^{-24}$.  
Since the kilonova occurred at a distance of $40 \Mpc$, possibly a nearby QN-NS collision could be brighter, but even if such a collision were to occur only $10 \pc$ from the Earth, the radiation signal would still be 11 orders of magnitude dimmer than the kilonova and unlikely to be detectable.  

%=========
For the self-collision of quark nuggets, the collision events can happen at a location away from the galactic plane.  
Therefore, one could search for transient sources at high latitude.  
For a source at a distance of $1 \pc$ from the Earth, the power per unit area is around $4 \times 10^{-20}\,\mbox{W}/\mbox{m}^2$.  
Assuming a telescope angular resolution of $1^\circ \times 1^\circ = (\pi/180)^2 \ \mbox{sr}$ and assuming a mono-energetic spectrum, the frequency weighted spectral intensity is $\nu I_\nu \approx 1 \times 10^{-16}\, \mbox{W}/(\mbox{m}^2\cdot\mbox{sr})$, which is two or three orders of magnitude below the cosmic gamma-ray background~\cite{Hill:2018trh}. 
If the collision events happen in a nearby location or within our solar system, the generated transient radiation signal could be detected by some radio, X-ray or gamma-ray telescopes.

%==================================
% Conclusion
%==================================
\section{Discussion and Conclusions}
\label{sec:Conclusion}

%=========
In this article we have studied an exotic form of Standard Model matter, called six-flavor quark matter. This new kind of matter can be formed through cosmological dynamics and exist in the universe today as nuggets, which are a candidate for dark matter.  
Our work is a natural extension of Witten's original work on three-flavor quark matter \cite{Witten:1984rs}, which assumed that Standard Model physics gives a first-order QCD phase transition (at $\mu_\B \approx 0$), and our work also builds upon several recent articles that study how beyond-the-SM physics can lead to a first order QCD phase transition \cite{Iso:2017uuu,Arunasalam:2017ajm,vonHarling:2017yew}.  
If a first-order quark-hadron phase transition occurs in the phase of unbroken electroweak symmetry (vanishing Higgs VEV), then our work demonstrates that nuggets of six-flavor quark matter will form.  

%=========
The properties of these 6FQM quark nuggets are summarized in \fref{fig:mass_radius}, which shows typical masses and radii of $M_\QN \sim 10^{7} - 10^{13} \gram$ and $R_\QN \sim 10^{-3} - 10^{-1} \cm$.  
We estimate the relic abundance of 6FQM nuggets, and the results are presented in \fref{fig:relic_abund}, which shows that these macroscopic dark matter candidate naturally explain the ratio $\Omega_\DM/\Omega_b \sim 5$.  
Due to the complicated nature of the first order QCD phase transition, it is challenging to make robust estimations, and whenever possible we have tried to be generous in our error estimates.  
In particular, the largest uncertainties in our calculations arise from estimating the hadronic-phase bubble nucleation rate, which affects the size and mass of the quarks nuggets through \erefs{eq:tgrow}{eq:Ri}; estimating the baryon-number ratio in the quark and hadronic phases, which affects the quark matter  relic abundance through \erefs{eq:r}{eq:Omega_ratio}; and estimating the thermal activation barrier height, which affects the quark nugget lifetime through \erefs{eq:tau}{eq:NB}.  

%=========
There are several directions in which our work could be further developed.  
Most notably, the formation of 6FQM nuggets requires the electroweak phase transition to be supercooled below the temperature of the QCD phase transition.  
Whereas we have presented a model that concretely implements this requirement in \sref{sec:EW_Trans}, the model is admittedly very tuned, and it would be useful to explore different implementations, perhaps in models that exhibit approximate scale invariance or shift-symmetric potentials.  
Similarly, such a scenario requires new physics coupled to the Higgs boson that may be testable at high-energy collider experiments.  
One may also want to consider whether heavy-ion colliders, which reach temperatures of $T \sim 200 \MeV$, may be able to produce the 6FQM phase, but the possibility seems remote since the $c$, $b$, and $t$ quarks are heavy today.  
Finally, it would be interesting to go beyond QCD quark nuggets to consider a confining hidden-sector with a first-order phase transition, providing a macroscopic dark matter candidate.

%=========
In conclusion, 6FQM nuggets provide an interesting candidate for dark matter with a unique set of observational signatures.  
Since observations do not currently constrain the presence of macroscopic dark matter in the 6FQM nugget mass window, one should view these dark matter candidates as a target for the next generation of gravitational, seismic, or astrophysical observations.  

%----------------------------------------------------------------
% Acknowledgements
%----------------------------------------------------------------
\subsubsection*{Acknowledgements}
We are grateful to several people for constructive comments and suggestions: Thomas Appelquist, Joshua Berger, Daniel Chung, Peter Cooper, Patrick Draper, Alex Drlica-Wagner, Joshua Frieman, Mark Hertzberg, Jonathan Kozaczuk, Manos Stamou, Glenn Starkman, and Lian-Tao Wang.  
The work of YB is supported by the U. S. Department of Energy under the contract DE-SC0017647.  
A.J.L. is supported at the University of Chicago by the Kavli Institute for Cosmological Physics through grant NSF PHY-1125897 and an endowment from the Kavli Foundation and its founder Fred Kavli.  

%----------------------------------------------------------------
% Appendix
%----------------------------------------------------------------
\begin{appendix}

%-------------------------------------------
\section{Higgs field profile around the 6FQM nuggets}
\label{sec:phi-profile}

\begin{figure}[t]
\begin{center}
\includegraphics[width=0.6\textwidth]{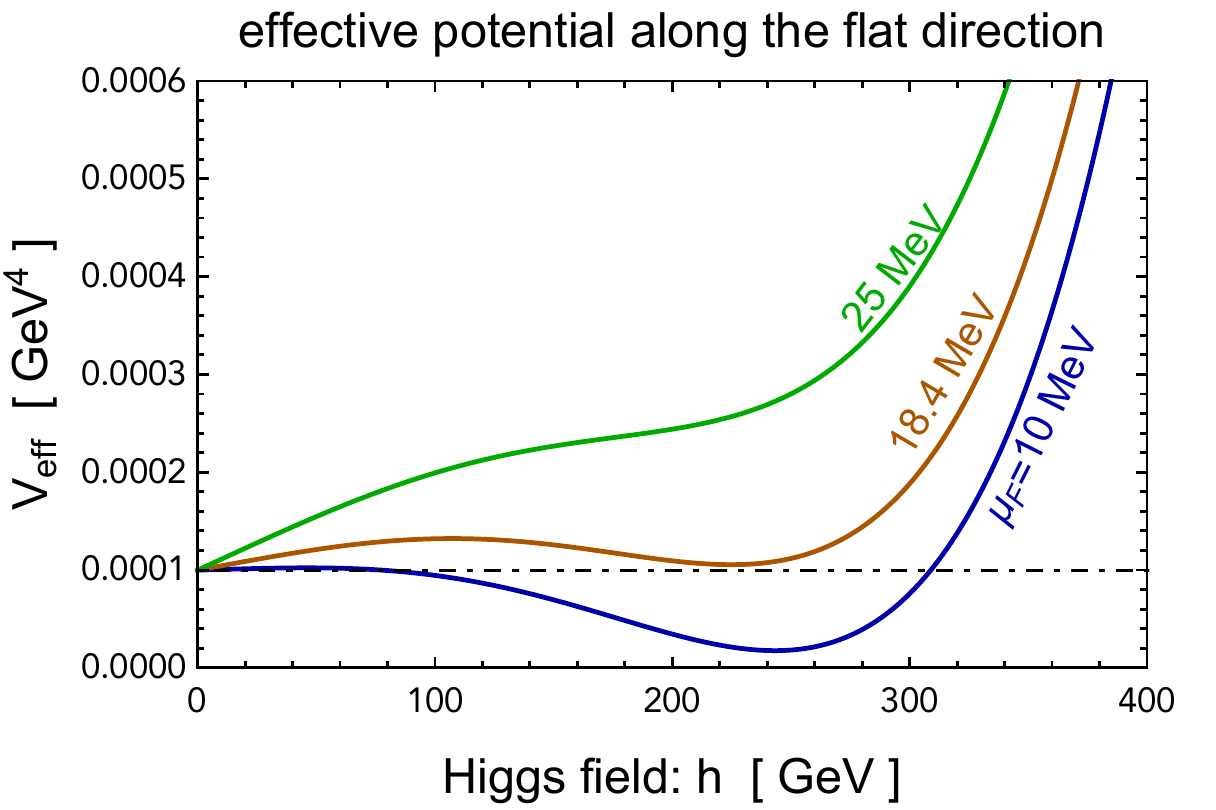}
\caption{
\label{fig:HiggsLeeWick}
The effective potential in \eqref{eq:leewick-potential} for different values of Fermi-momentum, $\mu_F$. The numerical values of the model parameters are the same as in Eq.~\eqref{eq:benchmark}.
}
\end{center}
\end{figure}

%=========
The Higgs field vanishes inside of the 6FQM nugget, $|\langle \Phi \rangle| = 0$, and it takes a nonzero value in the hadronic phase outside, $|\langle \Phi \rangle| = v_\EW$.  
One can calculate the profile of the Higgs field at the phase boundary from the effective potential, which sums the vacuum potential energy with the matter-induced energy.  
A similar calculation for Lee-Wick matter appears in Refs.~\cite{Lee:1974ma,Lee:1974uu}.  
The matter-induced potential is dominated by the top quark contribution, which has the largest Yukawa coupling $y_t \approx 1$, and the effective potential is calculated as 
\begin{align}\label{eq:leewick-potential}
	V_{\rm eff}(h)  = V(h) + 6 \int_F \frac{\ud ^3\pvec}{(2\pi)^3} \ \sqrt{\pvec^2 + \frac{1}{2}\,y_t^2\, h^2 } 
	\per
\end{align}
For degenerate Fermi matter, the momentum integral is cut off at the Fermi momentum, $|\pvec| \leq \mu_F$.  
Here, the $V(h)$ is taken to be the effective potential of $V(h, \phi)$ in \eref{eq:CW-potential} along the flat direction $\phi = \sqrt{ -2 \lambda_h / \lambda_{\rm mix}} \, h$.  
We show the effective potential in \fref{fig:HiggsLeeWick} as a function of the Higgs field for a few values of $\mu_F$. 

%=========

%=========
In a low-density region where $\mu_F$ is small, the energetically-preferred Higgs field value is the VEV $h = 246 \GeV$, but in a high-density region where $\mu_F$ is larger, it is $h=0$ that minimizes $V_{\rm eff}$ and electroweak symmetry restoration is preferred.  
For the benchmark model parameter point in \eref{eq:benchmark}, the transition value is around, $\mu_F^c = 18.4 \MeV$.  
Since $\mu \approx 100 \MeV$ for 6FQM [see \eref{eq:equilibrium}], the electroweak symmetry stays unbroken inside the 6FQM nugget, even though the outside world has $v_{\EW} = 246 \GeV$.  
At the phase boundary, the Higgs field profile is anticipated to vary (abruptly) from $0$ inside to $v_{\EW}$ on a length scale  that is small compared to the nugget's radius $R_\QN$.  

\end{appendix}

%----------------------------------------------------------------
% References
%----------------------------------------------------------------
%\bibliographystyle{JHEP}
%\bibliography{refs--quark_matter}

\providecommand{\href}[2]{#2}\begingroup\raggedright\begin{thebibliography}{}

\bibitem{Carr:1974nx}
Carr, Bernard J. and Hawking, S. W., \emph{{Black holes in the early
  Universe}}, {\emph{Mon. Not. Roy. Astron. Soc.} {\bfseries 168} (1974)
  399--415}.

\bibitem{Carr:2017jsz}
Carr, Bernard and Raidal, Martti and Tenkanen, Tommi and Vaskonen, Ville and
  Veerm{\"a}e, Hardi, \emph{{Primordial black hole constraints for extended
  mass functions}},
  \href{http://dx.doi.org/10.1103/PhysRevD.96.023514}{\emph{Phys. Rev.}
  {\bfseries D96} (2017) 023514},
  [\href{https://arxiv.org/abs/1705.05567}{{\ttfamily 1705.05567}}].

\bibitem{Witten:1984rs}
Witten, Edward, \emph{{Cosmic Separation of Phases}},
  \href{http://dx.doi.org/10.1103/PhysRevD.30.272}{\emph{Phys. Rev.} {\bfseries
  D30} (1984) 272--285}.

\bibitem{Farhi:1985ib}
Farhi, Edward and Jaffe, R. L., \emph{{Searching for Strange Matter by Heavy
  Ion Activation}},
  \href{http://dx.doi.org/10.1103/PhysRevD.32.2452}{\emph{Phys. Rev.}
  {\bfseries D32} (1985) 2452}.

\bibitem{Alcock:1985vc}
Alcock, Charles and Farhi, Edward, \emph{{The Evaporation of Strange Matter in
  the Early Universe}},
  \href{http://dx.doi.org/10.1103/PhysRevD.32.1273}{\emph{Phys. Rev.}
  {\bfseries D32} (1985) 1273}.

\bibitem{Olinto:1986je}
Olinto, Angela V., \emph{{On the Conversion of Neutron Stars Into Strange
  Stars}}, \href{http://dx.doi.org/10.1016/0370-2693(87)91144-0}{\emph{Phys.
  Lett.} {\bfseries B192} (1987) 71}.

\bibitem{Madsen:1986jg}
Madsen, J. and Heiselberg, H. and Riisager, K., \emph{{Does Strange Matter
  Evaporate in the Early Universe?}},
  \href{http://dx.doi.org/10.1103/PhysRevD.34.2947}{\emph{Phys. Rev.}
  {\bfseries D34} (1986) 2947--2955}.

\bibitem{Alcock:1986hz}
Alcock, Charles and Farhi, Edward and Olinto, Angela, \emph{{Strange stars}},
  \href{http://dx.doi.org/10.1086/164679}{\emph{Astrophys. J.} {\bfseries 310}
  (1986) 261--272}.

\bibitem{Alcock:1986bw}
Alcock, Charles and Farhi, Edward and Olinto, Angela, \emph{{STRANGE STARS}},
  \href{http://dx.doi.org/10.1016/0920-5632(91)90305-X}{\emph{Nucl. Phys. Proc.
  Suppl.} {\bfseries 24B} (1991) 93--102}.

\bibitem{Olinto:1987bt}
Olinto, Angela V., \emph{{Quark Matter in Astrophysics and Cosmology}},
  \href{http://dx.doi.org/10.1007/BF01574553}{\emph{Z. Phys.} {\bfseries C38}
  (1988) 303}.

\bibitem{Alcock:1988re}
Alcock, Charles and Olinto, Angela, \emph{{Exotic Phases of Hadronic Matter and
  their Astrophysical Application}},
  \href{http://dx.doi.org/10.1146/annurev.ns.38.120188.001113}{\emph{Ann. Rev.
  Nucl. Part. Sci.} {\bfseries 38} (1988) 161--184}.

\bibitem{Alcock:1988br}
Alcock, Charles and Olinto, Angela, \emph{{Evaporation of Strange Matter (And
  Similar Condensed Phases) at High Temperatures}},
  \href{http://dx.doi.org/10.1103/PhysRevD.39.1233}{\emph{Phys. Rev.}
  {\bfseries D39} (1989) 1233}.

\bibitem{Frieman:1989bu}
Frieman, Joshua A. and Olinto, Angela V., \emph{{Is the Submillisecond Pulsar
  Strange?}}, \href{http://dx.doi.org/10.1038/341633a0}{\emph{Nature}
  {\bfseries 341} (1989) 633--635}.

\bibitem{Olinto:1991rr}
Olinto, Angela V., \emph{{The Physics of strange matter}},  in \emph{{2nd
  International Workshop on Relativistic Aspects of Nuclear Physics Rio de
  Janeiro, Brazil, August 28-31, 1991}}, pp.~187--213, 1991.

\bibitem{Olinto:1991qq}
Olinto, A. V., \emph{{Converting neutron stars into strange stars}},
  \href{http://dx.doi.org/10.1016/0920-5632(91)90306-Y}{\emph{Nucl. Phys. Proc.
  Suppl.} {\bfseries 24B} (1991) 103--109}.

\bibitem{Olesen:1991zt}
Olesen, Michael L. and Madsen, Jes, \emph{{Boiling of strange quark matter}},
  \href{http://dx.doi.org/10.1103/PhysRevD.43.1069,
  10.1103/PhysRevD.44.566}{\emph{Phys. Rev.} {\bfseries D43} (1991)
  1069--1074}.

\bibitem{Olesen:1993iz}
Olesen, Michael L. and Madsen, Jes, \emph{{Boiling of strange quark matter in
  the early universe: strangeness conservation and nonequilibrium conditions}},
  \href{http://dx.doi.org/10.1103/PhysRevD.47.2313}{\emph{Phys. Rev.}
  {\bfseries D47} (1993) 2313--2323}.

\bibitem{Madsen:1998uh}
Madsen, Jes, \emph{{Physics and astrophysics of strange quark matter}},
  \href{http://dx.doi.org/10.1007/BFb0107314}{\emph{Lect. Notes Phys.}
  {\bfseries 516} (1999) 162--203},
  [\href{https://arxiv.org/abs/astro-ph/9809032}{{\ttfamily
  astro-ph/9809032}}].

\bibitem{Wilczek:2004rm}
Wilczek, Frank, \emph{{The Universe is a strange place}},
  \href{http://dx.doi.org/10.1016/j.nuclphysbps.2004.08.001}{\emph{Nucl. Phys.
  Proc. Suppl.} {\bfseries 134} (2004) 3--12},
  [\href{https://arxiv.org/abs/astro-ph/0401347}{{\ttfamily
  astro-ph/0401347}}].

\bibitem{Wilczek:2005ez}
Wilczek, Frank, \emph{{The Universe is a strange place}},
  \href{http://dx.doi.org/10.1142/S0217751X06032940}{\emph{Int. J. Mod. Phys.}
  {\bfseries A21} (2006) 2011--2025},
  [\href{https://arxiv.org/abs/physics/0511067}{{\ttfamily physics/0511067}}].

\bibitem{Madsen:2009tb}
Madsen, Jes, \emph{{Cavitation from bulk viscosity in neutron stars and strange
  stars}},  \href{https://arxiv.org/abs/0909.3724}{{\ttfamily 0909.3724}}.

\bibitem{Han:2009sj}
Han, Ke and Ashenfelter, Jeffrey and Chikanian, Alexei and Emmet, William and
  Finch, L. Evan and Heinz, Andreas and Madsen, Jes and Majka, Richard D. and
  Monreal, Benjamin and Sandweiss, Jack, \emph{{Search for stable Strange Quark
  Matter in lunar soil}},
  \href{http://dx.doi.org/10.1103/PhysRevLett.103.092302}{\emph{Phys. Rev.
  Lett.} {\bfseries 103} (2009) 092302},
  [\href{https://arxiv.org/abs/0903.5055}{{\ttfamily 0903.5055}}].

\bibitem{Lawson:2012zu}
Lawson, Kyle and Zhitnitsky, Ariel R., \emph{{Isotropic Radio Background from
  Quark Nugget Dark Matter}},
  \href{http://dx.doi.org/10.1016/j.physletb.2013.05.070}{\emph{Phys. Lett.}
  {\bfseries B724} (2013) 17--21},
  [\href{https://arxiv.org/abs/1210.2400}{{\ttfamily 1210.2400}}].

\bibitem{Fodor:2001pe}
Fodor, Z. and Katz, S. D., \emph{{Lattice determination of the critical point
  of QCD at finite T and mu}},
  \href{http://dx.doi.org/10.1088/1126-6708/2002/03/014}{\emph{JHEP} {\bfseries
  03} (2002) 014}, [\href{https://arxiv.org/abs/hep-lat/0106002}{{\ttfamily
  hep-lat/0106002}}].

\bibitem{Pisarski:1983ms}
Pisarski, Robert D. and Wilczek, Frank, \emph{{Remarks on the Chiral Phase
  Transition in Chromodynamics}},
  \href{http://dx.doi.org/10.1103/PhysRevD.29.338}{\emph{Phys. Rev.} {\bfseries
  D29} (1984) 338--341}.

\bibitem{Brown:1990ev}
Brown, Frank R. and Butler, Frank P. and Chen, Hong and Christ, Norman H. and
  Dong, Zhi-hua and Schaffer, Wendy and Unger, Leo I. and Vaccarino,
  Alessandro, \emph{{On the existence of a phase transition for QCD with three
  light quarks}},
  \href{http://dx.doi.org/10.1103/PhysRevLett.65.2491}{\emph{Phys. Rev. Lett.}
  {\bfseries 65} (1990) 2491--2494}.

\bibitem{Kuzmin:1992up}
Kuzmin, V. A. and Shaposhnikov, M. E. and Tkachev, I. I., \emph{{Strong CP
  violation, electroweak baryogenesis, and axionic dark matter}},
  \href{http://dx.doi.org/10.1103/PhysRevD.45.466}{\emph{Phys. Rev.} {\bfseries
  D45} (1992) 466--475}.

\bibitem{Konstandin:2011dr}
Konstandin, Thomas and Servant, Geraldine, \emph{{Cosmological Consequences of
  Nearly Conformal Dynamics at the TeV scale}},
  \href{http://dx.doi.org/10.1088/1475-7516/2011/12/009}{\emph{JCAP} {\bfseries
  1112} (2011) 009}, [\href{https://arxiv.org/abs/1104.4791}{{\ttfamily
  1104.4791}}].

\bibitem{Servant:2014bla}
Servant, Geraldine, \emph{{Baryogenesis from Strong $CP$ Violation and the QCD
  Axion}}, \href{http://dx.doi.org/10.1103/PhysRevLett.113.171803}{\emph{Phys.
  Rev. Lett.} {\bfseries 113} (2014) 171803},
  [\href{https://arxiv.org/abs/1407.0030}{{\ttfamily 1407.0030}}].

\bibitem{Arunasalam:2017ajm}
Arunasalam, Suntharan and Kobakhidze, Archil and Lagger, Cyril and Liang,
  Shelley and Zhou, Albert, \emph{{Low temperature electroweak phase transition
  in the Standard Model with hidden scale invariance}},
  \href{https://arxiv.org/abs/1709.10322}{{\ttfamily 1709.10322}}.

\bibitem{Iso:2017uuu}
Iso, Satoshi and Serpico, Pasquale D. and Shimada, Kengo,
  \emph{{QCD-Electroweak First-Order Phase Transition in a Supercooled
  Universe}},  \href{https://arxiv.org/abs/1704.04955}{{\ttfamily 1704.04955}}.

\bibitem{vonHarling:2017yew}
von Harling, Benedict and Servant, Geraldine, \emph{{QCD-induced Electroweak
  Phase Transition}},  \href{https://arxiv.org/abs/1711.11554}{{\ttfamily
  1711.11554}}.

\bibitem{Quigg:2009xr}
Quigg, Chris and Shrock, Robert, \emph{{Gedanken Worlds without Higgs:
  QCD-Induced Electroweak Symmetry Breaking}},
  \href{http://dx.doi.org/10.1103/PhysRevD.79.096002}{\emph{Phys. Rev.}
  {\bfseries D79} (2009) 096002},
  [\href{https://arxiv.org/abs/0901.3958}{{\ttfamily 0901.3958}}].

\bibitem{Farrar:2002ic}
Farrar, G. R., \emph{{A stable H dibaryon: Dark matter candidate within QCD?}},
  \href{http://dx.doi.org/10.1023/A:1025702431127}{\emph{Int. J. Theor. Phys.}
  {\bfseries 42} (2003) 1211--1218}.

\bibitem{Farrar:2017eqq}
Farrar, Glennys R., \emph{{Stable Sexaquark}},
  \href{https://arxiv.org/abs/1708.08951}{{\ttfamily 1708.08951}}.

\bibitem{Gross:2018ivp}
Gross, Christian and Polosa, Antonello and Strumia, Alessandro and Urbano,
  Alfredo and Xue, Wei, \emph{{Dark Matter in the Standard Model?}},
  \href{https://arxiv.org/abs/1803.10242}{{\ttfamily 1803.10242}}.

\bibitem{Zhitnitsky:2002qa}
Zhitnitsky, Ariel R., \emph{{'Nonbaryonic' dark matter as baryonic color
  superconductor}},
  \href{http://dx.doi.org/10.1088/1475-7516/2003/10/010}{\emph{JCAP} {\bfseries
  0310} (2003) 010}, [\href{https://arxiv.org/abs/hep-ph/0202161}{{\ttfamily
  hep-ph/0202161}}].

\bibitem{Oaknin:2003uv}
Oaknin, David H. and Zhitnitsky, Ariel, \emph{{Baryon asymmetry, dark matter
  and quantum chromodynamics}},
  \href{http://dx.doi.org/10.1103/PhysRevD.71.023519}{\emph{Phys. Rev.}
  {\bfseries D71} (2005) 023519},
  [\href{https://arxiv.org/abs/hep-ph/0309086}{{\ttfamily hep-ph/0309086}}].

\bibitem{Lawson:2013bya}
Lawson, Kyle and Zhitnitsky, Ariel R., \emph{{Quark (Anti) Nugget Dark
  Matter}},  in \emph{{Cosmic Frontier Workshop: Snowmass 2013 Menlo Park, USA,
  March 6-8, 2013}}, 2013.
\newblock \href{https://arxiv.org/abs/1305.6318}{{\ttfamily 1305.6318}}.

\bibitem{Liang:2016tqc}
Liang, Xunyu and Zhitnitsky, Ariel, \emph{{Axion field and the quark nugget's
  formation at the QCD phase transition}},
  \href{http://dx.doi.org/10.1103/PhysRevD.94.083502}{\emph{Phys. Rev.}
  {\bfseries D94} (2016) 083502},
  [\href{https://arxiv.org/abs/1606.00435}{{\ttfamily 1606.00435}}].

\bibitem{Ge:2017idw}
Ge, Shuailiang and Liang, Xunyu and Zhitnitsky, Ariel, \emph{{Cosmological
  axion and a quark nugget dark matter model}},
  \href{http://dx.doi.org/10.1103/PhysRevD.97.043008}{\emph{Phys. Rev.}
  {\bfseries D97} (2018) 043008},
  [\href{https://arxiv.org/abs/1711.06271}{{\ttfamily 1711.06271}}].

\bibitem{Arnold:1996dy}
Arnold, Peter Brockway and Son, Dam and Yaffe, Laurence G., \emph{{The Hot
  baryon violation rate is O (alpha-w**5 T**4)}},
  \href{http://dx.doi.org/10.1103/PhysRevD.55.6264}{\emph{Phys.Rev.} {\bfseries
  D55} (1997) 6264--6273},
  [\href{https://arxiv.org/abs/hep-ph/9609481}{{\ttfamily hep-ph/9609481}}].

\bibitem{Farhi:1984qu}
Farhi, Edward and Jaffe, R. L., \emph{{Strange Matter}},
  \href{http://dx.doi.org/10.1103/PhysRevD.30.2379}{\emph{Phys. Rev.}
  {\bfseries D30} (1984) 2379}.

\bibitem{Hogan:1984hx}
Hogan, C. J., \emph{{Nucleation of Cosmological Phase Transitions}},
  \href{http://dx.doi.org/10.1016/0370-2693(83)90553-1}{\emph{Phys. Lett.}
  {\bfseries 133B} (1983) 172--176}.

\bibitem{Kajantie:1986hq}
Kajantie, K. and Kurki-Suonio, H., \emph{{Bubble Growth and Droplet Decay in
  the Quark Hadron Phase Transition in the Early Universe}},
  \href{http://dx.doi.org/10.1103/PhysRevD.34.1719}{\emph{Phys. Rev.}
  {\bfseries D34} (1986) 1719--1738}.

\bibitem{Fodor:2001au}
Fodor, Z. and Katz, S. D., \emph{{A New method to study lattice QCD at finite
  temperature and chemical potential}},
  \href{http://dx.doi.org/10.1016/S0370-2693(02)01583-6}{\emph{Phys. Lett.}
  {\bfseries B534} (2002) 87--92},
  [\href{https://arxiv.org/abs/hep-lat/0104001}{{\ttfamily hep-lat/0104001}}].

\bibitem{Gasser:1986vb}
Gasser, J. and Leutwyler, H., \emph{{Light Quarks at Low Temperatures}},
  \href{http://dx.doi.org/10.1016/0370-2693(87)90492-8}{\emph{Phys. Lett.}
  {\bfseries B184} (1987) 83--88}.

\bibitem{DOnofrio:2015mpa}
D'Onofrio, Michela and Rummukainen, Kari, \emph{{Standard model cross-over on
  the lattice}},
  \href{http://dx.doi.org/10.1103/PhysRevD.93.025003}{\emph{Phys. Rev.}
  {\bfseries D93} (2016) 025003},
  [\href{https://arxiv.org/abs/1508.07161}{{\ttfamily 1508.07161}}].

\bibitem{Cohen:1993nk}
Cohen, Andrew G. and Kaplan, D. B. and Nelson, A. E., \emph{{Progress in
  electroweak baryogenesis}},
  \href{http://dx.doi.org/10.1146/annurev.ns.43.120193.000331}{\emph{Ann. Rev.
  Nucl. Part. Sci.} {\bfseries 43} (1993) 27--70},
  [\href{https://arxiv.org/abs/hep-ph/9302210}{{\ttfamily hep-ph/9302210}}].

\bibitem{Carena:1996wj}
Carena, Marcela and Quiros, M. and Wagner, C. E. M., \emph{{Opening the window
  for electroweak baryogenesis}},
  \href{http://dx.doi.org/10.1016/0370-2693(96)00475-3}{\emph{Phys. Lett.}
  {\bfseries B380} (1996) 81--91},
  [\href{https://arxiv.org/abs/hep-ph/9603420}{{\ttfamily hep-ph/9603420}}].

\bibitem{Bardeen:1995kv}
Bardeen, William A., \emph{{On naturalness in the standard model}},  in
  \emph{{Ontake Summer Institute on Particle Physics Ontake Mountain, Japan,
  August 27-September 2, 1995}}, 1995.

\bibitem{Tavares:2013dga}
Marques Tavares, Gustavo and Schmaltz, Martin and Skiba, Witold, \emph{{Higgs
  mass naturalness and scale invariance in the UV}},
  \href{http://dx.doi.org/10.1103/PhysRevD.89.015009}{\emph{Phys. Rev.}
  {\bfseries D89} (2014) 015009},
  [\href{https://arxiv.org/abs/1308.0025}{{\ttfamily 1308.0025}}].

\bibitem{Appelquist:2010xv}
{\scshape LSD} collaboration, Appelquist, Thomas and others, \emph{{Parity
  Doubling and the S Parameter Below the Conformal Window}},
  \href{http://dx.doi.org/10.1103/PhysRevLett.106.231601}{\emph{Phys. Rev.
  Lett.} {\bfseries 106} (2011) 231601},
  [\href{https://arxiv.org/abs/1009.5967}{{\ttfamily 1009.5967}}].

\bibitem{Appelquist:2014zsa}
{\scshape LSD} collaboration, Appelquist, T. and others, \emph{{Lattice
  simulations with eight flavors of domain wall fermions in SU(3) gauge
  theory}}, \href{http://dx.doi.org/10.1103/PhysRevD.90.114502}{\emph{Phys.
  Rev.} {\bfseries D90} (2014) 114502},
  [\href{https://arxiv.org/abs/1405.4752}{{\ttfamily 1405.4752}}].

\bibitem{Ade:2015xua}
{\scshape Planck} collaboration, Ade, P. A. R. and others, \emph{{Planck 2015
  results. XIII. Cosmological parameters}},
  \href{http://dx.doi.org/10.1051/0004-6361/201525830}{\emph{Astron.
  Astrophys.} {\bfseries 594} (2016) A13},
  [\href{https://arxiv.org/abs/1502.01589}{{\ttfamily 1502.01589}}].

\bibitem{Hasenfratz:1977dt}
Hasenfratz, Peter and Kuti, Julius, \emph{{The Quark Bag Model}},
  \href{http://dx.doi.org/10.1016/0370-1573(78)90076-5}{\emph{Phys. Rept.}
  {\bfseries 40} (1978) 75--179}.

\bibitem{Collins:1976yq}
Collins, John C. and Duncan, Anthony and Joglekar, Satish D., \emph{{Trace and
  Dilatation Anomalies in Gauge Theories}},
  \href{http://dx.doi.org/10.1103/PhysRevD.16.438}{\emph{Phys. Rev.} {\bfseries
  D16} (1977) 438--449}.

\bibitem{Nielsen:1977sy}
Nielsen, N. K., \emph{{The Energy Momentum Tensor in a Nonabelian Quark Gluon
  Theory}}, \href{http://dx.doi.org/10.1016/0550-3213(77)90040-2}{\emph{Nucl.
  Phys.} {\bfseries B120} (1977) 212--220}.

\bibitem{Appelquist:2010gy}
Appelquist, Thomas and Bai, Yang, \emph{{A Light Dilaton in Walking Gauge
  Theories}}, \href{http://dx.doi.org/10.1103/PhysRevD.82.071701}{\emph{Phys.
  Rev.} {\bfseries D82} (2010) 071701},
  [\href{https://arxiv.org/abs/1006.4375}{{\ttfamily 1006.4375}}].

\bibitem{Appelquist:2016viq}
Appelquist, Thomas and others, \emph{{Strongly interacting dynamics and the
  search for new physics at the LHC}},
  \href{http://dx.doi.org/10.1103/PhysRevD.93.114514}{\emph{Phys. Rev.}
  {\bfseries D93} (2016) 114514},
  [\href{https://arxiv.org/abs/1601.04027}{{\ttfamily 1601.04027}}].

\bibitem{Aoki:2014oha}
{\scshape LatKMI} collaboration, Aoki, Yasumichi and others, \emph{{Light
  composite scalar in eight-flavor QCD on the lattice}},
  \href{http://dx.doi.org/10.1103/PhysRevD.89.111502}{\emph{Phys. Rev.}
  {\bfseries D89} (2014) 111502},
  [\href{https://arxiv.org/abs/1403.5000}{{\ttfamily 1403.5000}}].

\bibitem{Aoki:2016wnc}
{\scshape LatKMI} collaboration, Aoki, Yasumichi and others, \emph{{Light
  flavor-singlet scalars and walking signals in $N_f=8$ QCD on the lattice}},
  \href{http://dx.doi.org/10.1103/PhysRevD.96.014508}{\emph{Phys. Rev.}
  {\bfseries D96} (2017) 014508},
  [\href{https://arxiv.org/abs/1610.07011}{{\ttfamily 1610.07011}}].

\bibitem{DeGrand:2015zxa}
DeGrand, Thomas, \emph{{Lattice tests of beyond Standard Model dynamics}},
  \href{http://dx.doi.org/10.1103/RevModPhys.88.015001}{\emph{Rev. Mod. Phys.}
  {\bfseries 88} (2016) 015001},
  [\href{https://arxiv.org/abs/1510.05018}{{\ttfamily 1510.05018}}].

\bibitem{Caprini:2015zlo}
Caprini, Chiara and others, \emph{{Science with the space-based interferometer
  eLISA. II: Gravitational waves from cosmological phase transitions}},
  \href{http://dx.doi.org/10.1088/1475-7516/2016/04/001}{\emph{JCAP} {\bfseries
  1604} (2016) 001}, [\href{https://arxiv.org/abs/1512.06239}{{\ttfamily
  1512.06239}}].

\bibitem{Caprini:2010xv}
Caprini, Chiara and Durrer, Ruth and Siemens, Xavier, \emph{{Detection of
  gravitational waves from the QCD phase transition with pulsar timing
  arrays}}, \href{http://dx.doi.org/10.1103/PhysRevD.82.063511}{\emph{Phys.
  Rev.} {\bfseries D82} (2010) 063511},
  [\href{https://arxiv.org/abs/1007.1218}{{\ttfamily 1007.1218}}].

\bibitem{Schettler:2010dp}
Schettler, Simon and Boeckel, Tillmann and Schaffner-Bielich, Jurgen,
  \emph{{Imprints of the QCD Phase Transition on the Spectrum of Gravitational
  Waves}}, \href{http://dx.doi.org/10.1103/PhysRevD.83.064030}{\emph{Phys.
  Rev.} {\bfseries D83} (2011) 064030},
  [\href{https://arxiv.org/abs/1010.4857}{{\ttfamily 1010.4857}}].

\bibitem{Aoki:2017aws}
Aoki, Mayumi and Goto, Hiromitsu and Kubo, Jisuke, \emph{{Gravitational Waves
  from Hidden QCD Phase Transition}},
  \href{http://dx.doi.org/10.1103/PhysRevD.96.075045}{\emph{Phys. Rev.}
  {\bfseries D96} (2017) 075045},
  [\href{https://arxiv.org/abs/1709.07572}{{\ttfamily 1709.07572}}].

\bibitem{Chen:2017cyc}
Chen, Yidian and Huang, Mei and Yan, Qi-Shu, \emph{{Gravitation waves from QCD
  and electroweak phase transitions}},
  \href{https://arxiv.org/abs/1712.03470}{{\ttfamily 1712.03470}}.

\bibitem{Ahmadvand:2017xrw}
Ahmadvand, M. and Bitaghsir Fadafan, K., \emph{{Gravitational waves generated
  from the cosmological QCD phase transition within AdS/QCD}},
  \href{http://dx.doi.org/10.1016/j.physletb.2017.07.039}{\emph{Phys. Lett.}
  {\bfseries B772} (2017) 747--751},
  [\href{https://arxiv.org/abs/1703.02801}{{\ttfamily 1703.02801}}].

\bibitem{Ahmadvand:2017tue}
Ahmadvand, M. and Bitaghsir Fadafan, K., \emph{{The cosmic QCD phase transition
  with dense matter and its gravitational waves from holography}},
  \href{http://dx.doi.org/10.1016/j.physletb.2018.01.066}{\emph{Phys. Lett.}
  {\bfseries B779} (2018) 1--8},
  [\href{https://arxiv.org/abs/1707.05068}{{\ttfamily 1707.05068}}].

\bibitem{Hindmarsh:2015qta}
Hindmarsh, Mark and Huber, Stephan J. and Rummukainen, Kari and Weir, David J.,
  \emph{{Numerical simulations of acoustically generated gravitational waves at
  a first order phase transition}},
  \href{http://dx.doi.org/10.1103/PhysRevD.92.123009}{\emph{Phys. Rev.}
  {\bfseries D92} (2015) 123009},
  [\href{https://arxiv.org/abs/1504.03291}{{\ttfamily 1504.03291}}].

\bibitem{Tisserand:2006zx}
{\scshape EROS-2} collaboration, Tisserand, P. and others, \emph{{Limits on the
  Macho Content of the Galactic Halo from the EROS-2 Survey of the Magellanic
  Clouds}}, \href{http://dx.doi.org/10.1051/0004-6361:20066017}{\emph{Astron.
  Astrophys.} {\bfseries 469} (2007) 387--404},
  [\href{https://arxiv.org/abs/astro-ph/0607207}{{\ttfamily
  astro-ph/0607207}}].

\bibitem{Niikura:2017zjd}
Niikura, Hiroko and Takada, Masahiro and Yasuda, Naoki and Lupton, Robert H.
  and Sumi, Takahiro and More, Surhud and More, Anupreeta and Oguri, Masamune
  and Chiba, Masashi, \emph{{Microlensing constraints on primordial black holes
  with the Subaru/HSC Andromeda observation}},
  \href{https://arxiv.org/abs/1701.02151}{{\ttfamily 1701.02151}}.

\bibitem{1992ApJ...386L...5G}
{Gould}, A., \emph{{Femtolensing of gamma-ray bursters}},
  \href{http://dx.doi.org/10.1086/186279}{\emph{Astrophysical Journal Letters}
  {\bfseries 386} (Feb., 1992) L5--L7}.

\bibitem{Barnacka:2012bm}
Barnacka, A. and Glicenstein, J. F. and Moderski, R., \emph{{New constraints on
  primordial black holes abundance from femtolensing of gamma-ray bursts}},
  \href{http://dx.doi.org/10.1103/PhysRevD.86.043001}{\emph{Phys. Rev.}
  {\bfseries D86} (2012) 043001},
  [\href{https://arxiv.org/abs/1204.2056}{{\ttfamily 1204.2056}}].

\bibitem{DeRujula:1984axn}
De Rujula, A. and Glashow, S. L., \emph{{Nuclearites: A Novel Form of Cosmic
  Radiation}}, \href{http://dx.doi.org/10.1038/312734a0}{\emph{Nature}
  {\bfseries 312} (1984) 734--737}.

\bibitem{Herrin:2005kb}
Herrin, Eugene T. and Rosenbaum, Doris C. and Teplitz, Vigdor L.,
  \emph{{Seismic search for strange quark nuggets}},
  \href{http://dx.doi.org/10.1103/PhysRevD.73.043511}{\emph{Phys. Rev.}
  {\bfseries D73} (2006) 043511},
  [\href{https://arxiv.org/abs/astro-ph/0505584}{{\ttfamily
  astro-ph/0505584}}].

\bibitem{Cyncynates:2016rij}
Cyncynates, David and Chiel, Joshua and Sidhu, Jagjit and Starkman, Glenn D.,
  \emph{{Reconsidering seismological constraints on the available parameter
  space of macroscopic dark matter}},
  \href{http://dx.doi.org/10.1103/PhysRevD.95.063006,
  10.1103/PhysRevD.95.129903}{\emph{Phys. Rev.} {\bfseries D95} (2017) 063006},
  [\href{https://arxiv.org/abs/1610.09680}{{\ttfamily 1610.09680}}].

\bibitem{Ackermann:2013zfa}
{\scshape Fermi-LAT} collaboration, Ajello, M. and others, \emph{{The First
  Fermi LAT Gamma-Ray Burst Catalog}},
  \href{http://dx.doi.org/10.1088/0067-0049/209/1/11}{\emph{Astrophys. J.
  Suppl.} {\bfseries 209} (2013) 11},
  [\href{https://arxiv.org/abs/1303.2908}{{\ttfamily 1303.2908}}].

\bibitem{vonKienlin:2014nza}
von Kienlin, Andreas and others, \emph{{The Second Fermi GBM Gamma-Ray Burst
  Catalog: The First Four Years}},
  \href{http://dx.doi.org/10.1088/0067-0049/211/1/13}{\emph{Astrophys. J.
  Suppl.} {\bfseries 211} (2014) 13},
  [\href{https://arxiv.org/abs/1401.5080}{{\ttfamily 1401.5080}}].

\bibitem{Price:1983ax}
Price, P. B. and Guo, Shi-lun and Ahlen, S. P. and Fleischer, R. L.,
  \emph{{Search for {GUT} Magnetic Monopoles at a Flux Level Below the Parker
  Limit}}, \href{http://dx.doi.org/10.1103/PhysRevLett.52.1265}{\emph{Phys.
  Rev. Lett.} {\bfseries 52} (1984) 1265}.

\bibitem{Jacobs:2014yca}
Jacobs, David M. and Starkman, Glenn D. and Lynn, Bryan W., \emph{{Macro Dark
  Matter}}, \href{http://dx.doi.org/10.1093/mnras/stv774}{\emph{Mon. Not. Roy.
  Astron. Soc.} {\bfseries 450} (2015) 3418--3430},
  [\href{https://arxiv.org/abs/1410.2236}{{\ttfamily 1410.2236}}].

\bibitem{Camenzind:2007}
Max Camenzind, \emph{Compact Objects in Astrophysics}.
\newblock Springer.

\bibitem{Hotokezaka:2012ze}
Hotokezaka, Kenta and Kiuchi, Kenta and Kyutoku, Koutarou and Okawa, Hirotada
  and Sekiguchi, Yu-ichiro and Shibata, Masaru and Taniguchi, Keisuke,
  \emph{{Mass ejection from the merger of binary neutron stars}},
  \href{http://dx.doi.org/10.1103/PhysRevD.87.024001}{\emph{Phys. Rev.}
  {\bfseries D87} (2013) 024001},
  [\href{https://arxiv.org/abs/1212.0905}{{\ttfamily 1212.0905}}].

\bibitem{TheLIGOScientific:2017qsa}
{\scshape Virgo, LIGO Scientific} collaboration, Abbott, B.P. and others,
  \emph{{GW170817: Observation of Gravitational Waves from a Binary Neutron
  Star Inspiral}},
  \href{http://dx.doi.org/10.1103/PhysRevLett.119.161101}{\emph{Phys. Rev.
  Lett.} {\bfseries 119} (2017) 161101},
  [\href{https://arxiv.org/abs/1710.05832}{{\ttfamily 1710.05832}}].

\bibitem{GBM:2017lvd}
Abbott, B. P. and others, \emph{{Multi-messenger Observations of a Binary
  Neutron Star Merger}},
  \href{http://dx.doi.org/10.3847/2041-8213/aa91c9}{\emph{Astrophys. J.}
  {\bfseries 848} (2017) L12},
  [\href{https://arxiv.org/abs/1710.05833}{{\ttfamily 1710.05833}}].

\bibitem{Monitor:2017mdv}
{\scshape Virgo, Fermi-GBM, INTEGRAL, LIGO Scientific} collaboration, Abbott,
  B. P. and others, \emph{{Gravitational Waves and Gamma-rays from a Binary
  Neutron Star Merger: GW170817 and GRB 170817A}},
  \href{http://dx.doi.org/10.3847/2041-8213/aa920c}{\emph{Astrophys. J.}
  {\bfseries 848} (2017) L13},
  [\href{https://arxiv.org/abs/1710.05834}{{\ttfamily 1710.05834}}].

\bibitem{Hill:2018trh}
Hill, Ryley and Masui, Kiyoshi W. and Scott, Douglas, \emph{{The Spectrum of
  the Universe}},  \href{https://arxiv.org/abs/1802.03694}{{\ttfamily
  1802.03694}}.

\bibitem{Lee:1974ma}
Lee, T. D. and Wick, G. C., \emph{{Vacuum Stability and Vacuum Excitation in a
  Spin 0 Field Theory}},
  \href{http://dx.doi.org/10.1103/PhysRevD.9.2291}{\emph{Phys. Rev.} {\bfseries
  D9} (1974) 2291--2316}.

\bibitem{Lee:1974uu}
Lee, T. D. and Margulies, M., \emph{{Interaction of a Dense Fermion Medium with
  a Scalar Meson Field}}, \href{http://dx.doi.org/10.1103/PhysRevD.12.4008,
  10.1103/PhysRevD.11.1591}{\emph{Phys. Rev.} {\bfseries D11} (1975) 1591}.

\end{thebibliography}\endgroup

\providecommand{\href}[2]{#2}\begingroup\raggedright\endgroup

\end{document}